\documentclass[5p,twocolumn]{elsarticle} %

\usepackage[T1]{fontenc}
\usepackage{lmodern}
\usepackage[normalem]{ulem}

\usepackage{epstopdf} %
\usepackage{hyperref}
\usepackage{xcolor}
\hypersetup{
    colorlinks=true,
    linkcolor=blue,
    filecolor=magenta,      
    urlcolor=cyan,
}
\usepackage[caption=false]{subfig} %
\usepackage{siunitx}

\usepackage{libertine}
\usepackage[libertine]{newtxmath}

\journal{ArXiv}
\newcommand{\supplementary}{Supplementary Material}

\begin{document}

\begin{frontmatter}

\title{Activation entropy of dislocation glide in body-centered cubic metals from atomistic simulations}

\author[label1,label2,label3,label5]{Arnaud Allera\corref{cor1}}
\ead{arnaud.allera@asnr.fr}
\author[label4]{Thomas D. Swinburne}
\author[label3]{Alexandra M. Goryaeva}
\author[label3,label6]{Baptiste Bienvenu}
\author[label1]{Fabienne Ribeiro}
\author[label5]{Michel Perez}
\author[label3]{Mihai-Cosmin Marinica}
\author[label2]{David Rodney\corref{cor1}}
\ead{david.rodney@univ-lyon1.fr}

\cortext[cor1]{Corresponding author}

\address[label1]{ASNR/PSN-RES/SEMIA/LSMA Centre d'études de Cadarache, F-13115 Saint Paul-lez-Durance, France}
\address[label2]{Univ. Lyon, UCBL, Institut Lumière Matière, UMR CNRS 5306, F-69622 Villeurbanne, France}
\address[label3]{Univ. Paris-Saclay, CEA, Service de Recherche en Corrosion et Comportement des Matériaux, SRMP, F-91191, Gif-sur-Yvette, France}
\address[label5]{Univ. Lyon, INSA Lyon, UCBL, MATEIS, UMR CNRS 5510, F-69621 Villeurbanne, France}
\address[label4]{Aix-Marseille Univ., CINaM, UMR CNRS 7325, Campus de Luminy, F-13288 Marseille, France}
\address[label6]{Max Planck Institute for Sustainable Materials, Max-Planck-Straße 1, 40237, Düsseldorf, Germany
}

\begin{abstract}

\noindent The activation entropy of dislocation glide, a key process controlling the strength of many metals, is often assumed to be constant or linked to enthalpy through the empirical Meyer-Neldel law—both of which are simplified approximations. In this study, we take a more direct approach by calculating the activation Gibbs energy for kink-pair nucleation on screw dislocations of two body-centered cubic metals, iron and tungsten. To ensure reliability, we develop machine learning interatomic potentials for both metals, carefully trained on dislocation data from density functional theory. Our findings reveal that dislocations undergo harmonic transitions between Peierls valleys, with an activation entropy that remains largely constant, regardless of temperature or applied stress. We use these results to parameterize a thermally-activated model of yield stress, which consistently matches experimental data in both iron and tungsten. Our work challenges recent studies using classical potentials, which report highly varying activation entropies, and suggests that simulations relying on classical potentials—widely used in materials modeling—could be significantly influenced by overestimated entropic effects.
\end{abstract}

\end{frontmatter}

\section{Introduction}

A classical thermally-activated process in metallurgy is the glide of screw dislocations in body-centered cubic (BCC) metals~\cite{caillard2003thermally}. Despite its importance, this process has
so far been primarily treated phenomenologically~\cite{po2016phenomenological}, assuming a simple expression for the activation entropy and a linear relation between the applied stress tensor and the critical resolved shear stress at 0K. In particular, the interplay between applied stress and temperature, i.e.~the stress- and temperature-dependence of the Gibbs activation energy for dislocation glide, remains a theoretical obstacle where assumptions are commonly used without appropriate justification.

\begin{figure}
    \centering
    \includegraphics[width=0.8\columnwidth]{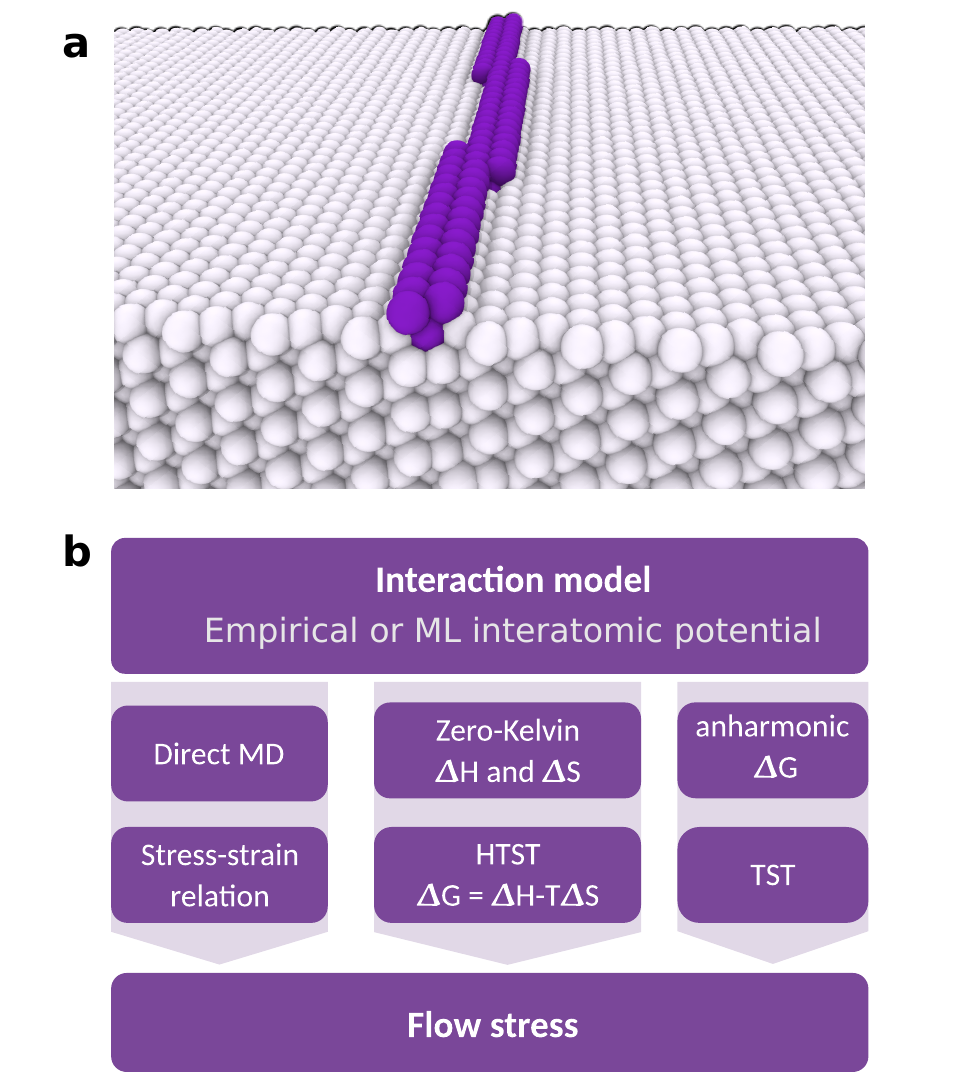}
    \caption{\textbf{Overview of dislocation glide simulations and associated computational methods.} 
    (a) A screw dislocation migrating from one Peierls valley to the next by a kink-pair mechanism. The 3D atomic structure was rendered with Ovito \cite{stukowski2009visualization}. The upper part of the crystal is not shown for clarity.
    (b) Overview of different computational approaches to predict the flow stress.
    }
    \label{fig:f1}
\end{figure}

Dislocation glide can be simulated by direct molecular dynamics (MD) simulations~\cite{domain2005simulation,chaussidon2006glide,zepeda2017probing} (as illustrated in Fig.~\ref{fig:f1} (a) and left track in Fig.~\ref{fig:f1} (b)),
but only at high temperatures and/or high strain rates.
To access more general conditions, one can use the Transition State Theory (TST) \cite{laidler1983development}, which expresses the average forward dislocation velocity $v$ at a temperature $T$ under an applied stress tensor $[\sigma]$:

\begin{equation}
    v = \nu L \exp \Big(-\frac{ \Delta G([\sigma],T)}{\mathrm{k_B} T } \Big),
    \label{eq:velo}
\end{equation}
with $\nu$ an attempt frequency, $L$ the dislocation length, $\mathrm{k_B} T$ the thermal energy and $\Delta G$ the Gibbs activation energy for kink-pair nucleation. In BCC metals, the activation free enthalpy, $\Delta G$, depends a priori on all components and sign of the applied stress tensor due to non-Schmid effects~\cite{dezerald2016plastic,kraych2019non}. However, in this work, we will only focus on the effect of a resolved shear stress applied in the dislocation glide plane. 

Due to the difficulty in computing Gibbs activation energies, \textit{ad hoc} simplifications are often used. The most classical is the harmonic TST (HTST)~\cite{vineyard1957frequency} (see central track in Fig.~\ref{fig:f1} (b)). 
Both the activation entropy $\Delta S_h$ and enthalpy $\Delta H_h$ are then independent of $T$. $\Delta S_h$ is computed by diagonalizing the Hessian matrix of the system in its zero-Kelvin initial state (a straight screw dislocation in a Peierls valley) and activated state (an unstable double-kinked dislocation in-between Peierls valleys, illustrated in Fig.~\ref{fig:f1} (a))~\cite{proville2012quantum}.
However, the diagonalizations are very demanding in computational resources for systems containing the $N \sim 10^5$ atoms needed to model dislocations \cite{proville2012quantum}. 
As a result, further approximations are commonly applied.  
The most common is to simply neglect the stress dependence of $\Delta S_h$, which becomes a constant that can be absorbed into the attempt frequency $\nu^* = \nu \exp(\Delta S_h/\mathrm{k_B})$.  
However, there is evidence that activation entropies in dislocation-related processes may exhibit a stress dependence, including dislocation nucleation~\cite{warner2009origins,ryu2011entropic}, obstacle bypass~\cite{saroukhani2016harnessing}, cross-slip~\cite{esteban2020influence}, and the glide of straight dislocations~\cite{gilbert2013free}.  
It is important to note here that these calculations were performed using empirical interatomic potentials. 
An alternative approach to incorporate stress dependence without explicitly computing $\Delta S$ is to invoke the Meyer-Neldel (MN) law, or compensation rule~\cite{yelon2006multi}, which states that the activation entropy of a process is proportional to its activation enthalpy,  $\Delta S = \Delta H/T_{\textrm{MN}}$, where $T_{\textrm{MN}}$ is a characteristic temperature. 
Despite its widespread use, the MN law lacks a firm physical foundation, and counterexamples have been reported in both experiments~\cite{meiling1999inverse} and computational simulations~\cite{koziatek2013inverse,gelin2020enthalpy,zotov2022entropy,wang2023stress}.  
Here also, it should be emphasized that these computational studies were based on empirical potentials.

Empirical potentials, such as embedded atom method (EAM) potentials, have significantly advanced the modeling of BCC screw dislocations. However, most fail to accurately reproduce the potential energy landscape of screw dislocations obtained from density functional theory (DFT) calculations~\cite{dezerald2016plastic} and the temperature dependence of elastic constants compared to experimental data~\cite{zhong2023anharmonic}.  
Machine-learning interatomic potentials (MLIPs) provide a promising alternative to overcome these limitations~\cite{deringer2019machine} at a computational cost that has now become manageable.  
However, while some recently developed MLIPs for Fe have been explicitly trained on dislocation configurations~\cite{byggmastar2022multiscale,goryaeva2021efficient,meng2021general}, their datasets remain limited and often focus primarily on zero-Kelvin properties.

In this work, we investigate the influence of vibrational entropy on dislocation mobility without relying on a priori simplifications, using state-of-the-art interatomic potentials.
We modeled isolated dislocations in BCC metals and directly computed their (possibly anharmonic) Gibbs activation energy for kink-pair nucleation using the linear-scaling projected average force integrator (PAFI)\cite{swinburne2018unsupervised}.
This offers an alternative computational route to estimate the flow stress, following the path highlighted in Fig.\ref{fig:f1}(b).
To accurately capture activation thermodynamics, we developed MLIPs for two common BCC metals, Fe and W, by refining the models from Ref.\cite{goryaeva2021efficient} and expanding the training dataset to include a wider variety of dislocation configurations.
Further details on the simulation setup, computational methods, and MLIPs are provided in the Methods section.

\section{Results}

\subsection{Machine-learning interatomic potentials}
\label{sec:MLIP}

\begin{figure*}
    \centering
    \includegraphics[width=\textwidth]{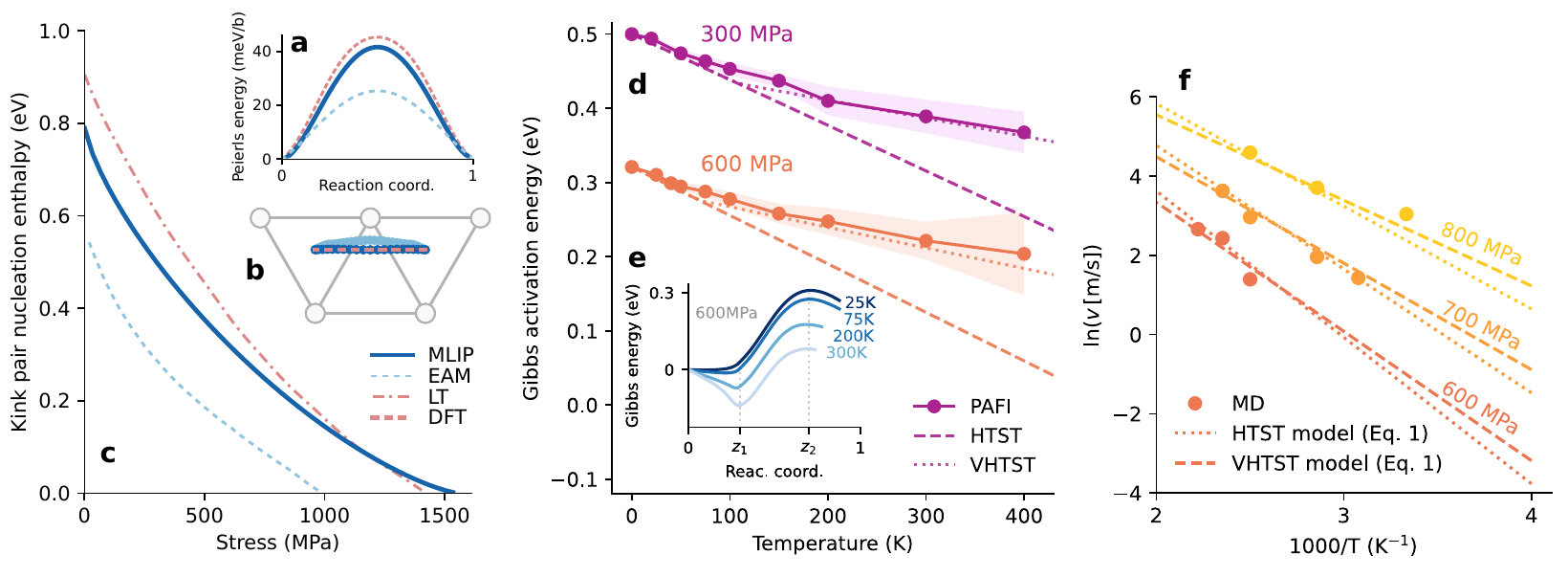}
    \caption{
    \textbf{{MLIP predicts a harmonic behavior and a stress-independent activation entropy.}}
    (a) Peierls barrier for a ½[111] straight screw dislocation and (b) corresponding core trajectory, computed with the present MLIP for Fe and a classical EAM potential \cite{proville2012quantum}, compared to  DFT data~\cite{bienvenu2022ab}. 
    (c) Kink-pair nucleation enthalpy computed with the EAM and MLIP, compared to a line tension (LT) model parameterized on DFT calculations~\cite{dezerald2015first}. 
    (d) Gibbs activation energy for kink-pair nucleation computed. PAFI calculations are compared to variational and non-variational HTST predictions. Shaded regions represent the error estimation of the PAFI method, as described in \supplementary{}.
    (e) Examples of Gibbs energy profiles at \SI{600}{\mega\pascal} showing a shift of the minimum Gibbs energy configuration.  
    (f) Dislocation velocity computed by direct MD simulations with the MLIP potential (circles) under an constant applied shear stress, compared to Eq.~\ref{eq:velo} parameterized with data from (c-d) with $\nu$= \SI{3.8e9}{\hertz} for HTST (dotted lines) and \SI{9.2e10}{\hertz} for variational HTST (dashed lines).}
    \label{fig:4}
\end{figure*}

We begin by examining the Gibbs activation energy for kink-pair nucleation in Fe. To this end, we developed a MLIP for Fe, building upon the framework established in Ref.~\cite{goryaeva2021efficient} and successfully applied to free energy calculations of defects in BCC metals~\cite{zhong2023anharmonic,zhong2025unraveling}. 
The training dataset was significantly expanded to include a wide range of dislocation configurations. Notably, in addition to the Peierls barrier, the dataset incorporates the hard-to-split path~\cite{dezerald2016plastic} and single-kink configurations~\cite{ventelon2009atomistic}, which are considered for the first time.
These configurations were evaluated both at \SI{0}{\kelvin} and finite temperatures, with non-equilibrium states sampled using constrained MD simulations (see \supplementary{} for details). The resulting MLIP accurately reproduces the Peierls mechanism, core trajectory, and kink-pair nucleation enthalpy (see Fig.~\ref{fig:4}~(a-c)), as well as core eigenstrains (see \supplementary{}), demonstrating excellent agreement with DFT data. For comparison, we included in Fig.~\ref{fig:4} predictions of a classical EAM potential for Fe \cite{proville2012quantum}, which will be further used in Section \ref{sec:EAM}.

Using the MLIP, we employed the PAFI method to compute minimum Gibbs energy paths (MGEP) without any \textit{a priori} assumptions. The resulting Gibbs activation energies, obtained at \SI{300}{} and \SI{600}{\mega\pascal}, are shown in Fig.~\ref{fig:4}~(d). At low temperatures, the enthalpy decreases linearly with increasing temperature, indicating a harmonic regime that extends up to $T_0 \simeq \SI{100}{\kelvin}$. Beyond this threshold, the Gibbs activation energy deviates from the harmonic extrapolation and decreases more gradually.
The discrepancy between HTST and anharmonic calculations is further illustrated in Fig.~\ref{fig:4}~(e) and discussed in \supplementary{}. This difference primarily arises from a change in the Gibbs energy profile: above $T_0$, the position of the barrier minimum shifts from $z = 0$ to a finite coordinate $z_1$, while the barrier maximum remains mostly fixed at a coordinate $z_2$ (see Fig.~\ref{fig:4}(e)). Unfortunately, we could not identify any apparent structural changes that may account for the lowering of the Gibbs free energy in the finite-temperature configurations generated by PAFI.
In \supplementary{}, we detail a statistical outlier analysis of the shifted transition pathway sampled by PAFI, demonstrating that the sampled configurations remain in the interpolation domain of the MLIP.

To test the validity of the harmonic assumption along the transition path and extend the harmonic regime, we use the variational HTST (VHTST)~\cite{truhlar1984variational}, where the Gibbs energy barrier is computed harmonically between the configurations that maximize the barrier height, a method recently employed to model dislocation nucleation kinetics~\cite{bagchi2024anomalous}.
We thus approximate $\Delta G_{\textrm{VHTST}} = \Delta H_h(z_2) - T \Delta S_h(z_2)$ if $T<T_0$  and $\Delta G_{\textrm{VHTST}} = \Delta H_h(z_2)-\Delta H_h(z_1) - T (\Delta S_h(z_2)-\Delta S_h(z_1))$ if $T> T_0$, where $\Delta H_h(z)$ and $\Delta S_h(z)$ are the 0K enthalpy and entropy differences between the configuration at reaction coordinate $z$ and the initial configuration (at $z=0)$.
The result is reported in Fig~\ref{fig:4}~({d}) as dotted lines, showing a very good agreement with the anharmonic PAFI calculations over the entire temperature range considered here.

Perhaps the most striking result is the remarkably weak dependence of the activation entropy on the applied stress: at low temperature, in the harmonic regime, $\Delta S(z_2) = 6.3 \mathrm{k_B}$, while above $T_0$, $\Delta S(z_2) - \Delta S(z_1)=1.6\mathrm{k_B}$. In the harmonic regime, the activation entropy is thus independent of both temperature and applied stress, providing, for the first time, a direct justification for the commonly used simplification of a constant entropy.

To further validate our calculations, we conducted MD simulations with the MLIP for different resolved shear stresses and temperatures. The measured average dislocation velocities are presented in Fig.~\ref{fig:4}~(f) and compared to the predictions of both HTST, $v = \nu L \exp (-\Delta G_h/ \mathrm{k_B} T)$, and VHTST, $v = \nu L \exp (-\Delta G_{\textrm{VHTST}}/ \mathrm{k_B} T)$, where $\nu$ is the only fitting parameter.  
Both approximations yield nearly identical velocity predictions, as differences in entropy are absorbed into the fitted prefactor $\nu$, while the enthalpic change is negligible, given that $\Delta H_h(z_1) \ll \Delta H_h(z_2)$. The strong agreement between MD and theoretical predictions confirms the harmonic nature of the glide transition and the absence of significant stress dependence of the activation entropy for this MLIP.

To evaluate the generality of the activation enthalpy and entropy behavior observed in Fe, we developed a second MLIP for W. As in Fe, a database built in previous works \cite{goryaeva2021efficient,zhong2023anharmonic,zhong2024machine} was enriched with finite-temperature dislocation configurations.
The corresponding results, presented in \supplementary{}, confirm the existence of a harmonic regime up to at least 100 K with an activation entropy independent of the applied stress. %
Remarkably, despite the significant differences between Fe and W -such as their elastic constants - the harmonic activation entropy in W is found to be approximately $8 \mathrm{k_B}$, closely matching the value obtained for Fe. This unexpected similarity suggests a degree of universality in the contribution of vibrational entropy to kink pair nucleation across BCC metals.

\subsection{EAM Potentials}
\label{sec:EAM}

\begin{figure*}
    \centering
    \includegraphics[width=1\linewidth]{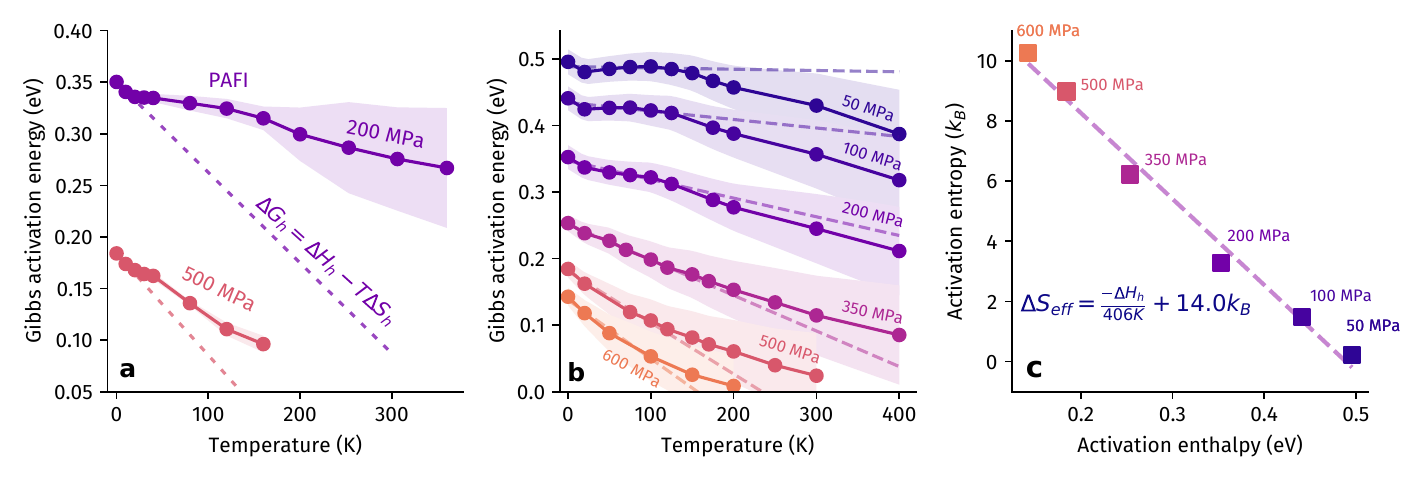}
    \caption{\textbf{Empirical potential predicts strong anharmonicity and an inverse Meyer-Neldel relation.} (a) Gibbs activation energy as a function of temperature at 200 and \SI{500}{\mega\pascal}, obtained with PAFI (circles and full lines) and the HTST approximation using Hessian matrix diagonalizations (dashed lines). 
    Shaded regions represent the error estimation of the PAFI method, as described in \supplementary{}.
    Lines are colored based on the associated stress value.
    (b) Gibbs activation energy as a function of temperature for a large range of stresses and temperatures. Linear fits up to \SI{100}{\kelvin} with slope $-\Delta S_{\mathrm{eff}}$ are shown as dashed lines. 
    Note that calculations presented in (a) were performed with a high number of independent samplings (ranging from 150 to 250), in order to accurately capture subtle variations of Gibbs activation energy at very low temperature.
    (c) Effective entropy $\Delta S_{\mathrm{eff}}$ as a function of enthalpy.
    }
    \label{fig:3}
\end{figure*}

The strikingly constant entropy obtained with the MLIPs for Fe and W contrasts with the results obtained in the literature using empirical potentials and mentioned in the Introduction. To verify this point, we used two EAM potentials, one for Fe \cite{proville2012quantum}, the other for W \cite{marinica2013interatomic}, widely used to simulate dislocation glide at the atomic scale. We computed their Gibbs activation energies for kink-pair nucleation, and since the calculations were much faster than with the MLIPs, we could consider a wide range of applied stresses. Fig.~\ref{fig:3} considers the case of Fe, while the results in W are shown in \supplementary{}. Fig.~\ref{fig:3}~(a) shows the Gibbs activation energies at \SI{200}{} and \SI{500}{\mega\pascal}.
We see that $\Delta G$ exhibits a temperature dependence more complex than with the MLIP, with a drastic departure from the harmonic prediction (dashed lines) above temperatures as low as \SI{20}{\kelvin}.
Note that we have checked that for the moderate stresses and temperatures considered here, classically discussed anharmonic effects, temperature dependence of the elastic moduli~\cite{ryu2011entropic} and thermal dilatation~\cite{wang2023stress} do not affect the energy barriers (see \supplementary{}).
The main effect is rather a widening of the activated kink pair \cite{swinburne2018unsupervised}, which reflects a deviation of the MGEPs at finite temperature, as detailed in \supplementary{}.

Although the HTST approximation appears inapplicable beyond about \SI{20}{\kelvin} with the present EAM potential, 
the evolution of $\Delta G$ shown in Fig.~\ref{fig:3}~(b) can still be approximated by a linear relation with reasonable accuracy below \SI{100}{\kelvin} (dashed lines in Fig.~\ref{fig:3}~(b)). 
This defines an effective temperature-independent entropy, $\Delta S_{\mathrm{eff}}(\tau)$.  
Interestingly, we see in Fig.~\ref{fig:3} (b) that the initial slope of $\Delta G$ becomes steeper at higher stresses, which means that $\Delta S_{\mathrm{eff}}(\tau)$ increases with the applied stress.
Plotted against $\Delta H_h(\tau)$ in Fig. \ref{fig:3} (c), we find a linear variation, which follows an inverse Meyer-Neldel law, or reinforcement effect: $\Delta S_{\mathrm{eff}}$ decreases with increasing $\Delta H_h$, instead of increasing as assumed by the MN compensation law. 
Note also that $\Delta S_{\mathrm{eff}}$ varies by about $10 \mathrm{k_B}$ between 0 and \SI{700}{\mega\pascal},  in stark contrast with the results obtained with the MLIPs. 
The emergence of an inverse Meyer-Neldel law is consistent with recent works presented in Refs.~\cite{zotov2022entropy,wang2023stress}, indicating that it might be a general trend of EAM potentials. This conclusion is reinforced by calculations performed with the EAM potential for W. As presented in \supplementary{}, the potential also predicts a harmonic regime limited to very low temperatures with again an inverse MN rule between the enthalpy and a strongly varying entropy.

\subsection{From Gibbs energy to yield stress}
\label{sec:ystress}

\begin{figure*}[htb]
    \centering
    \subfloat{\includegraphics[width=0.6\linewidth]{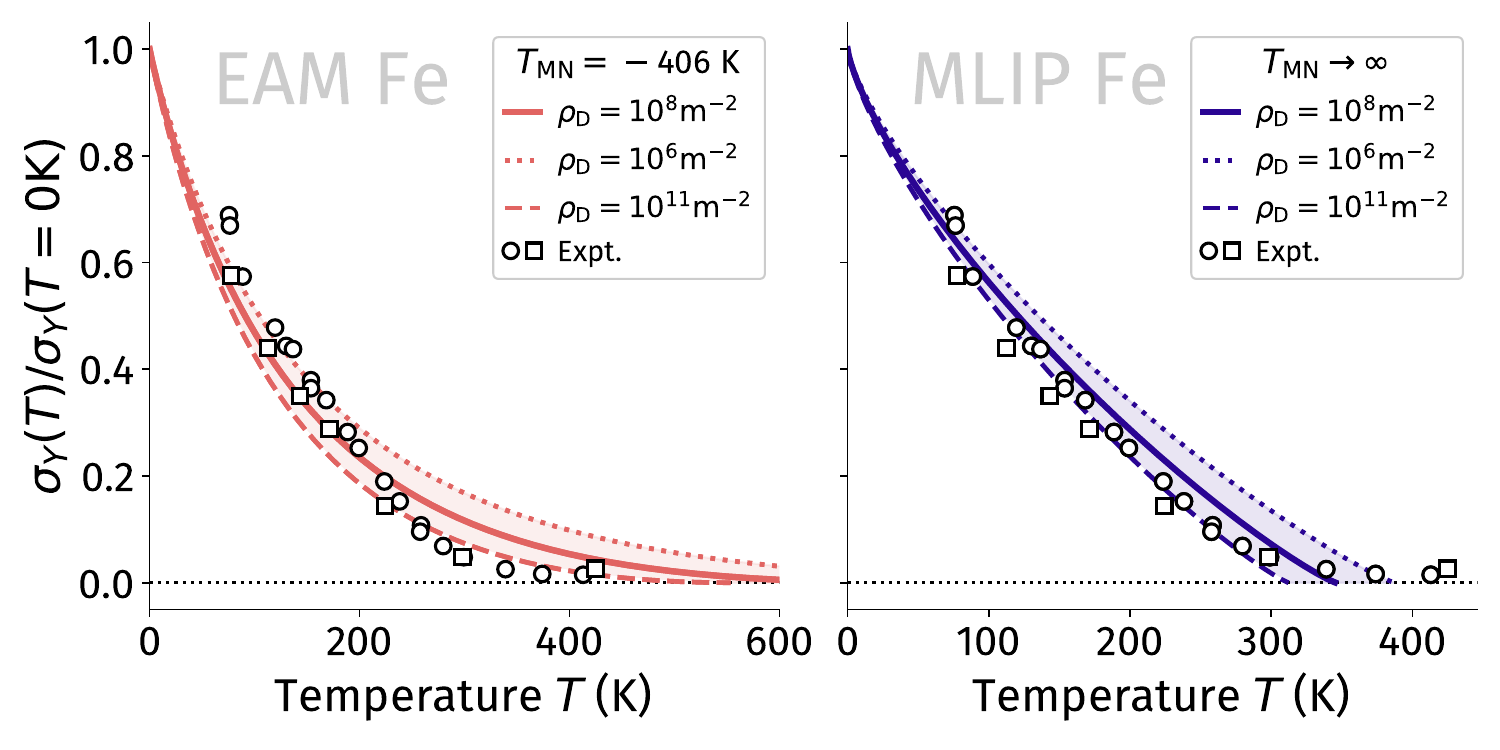}}\\
    \subfloat{\includegraphics[width=0.6\linewidth]{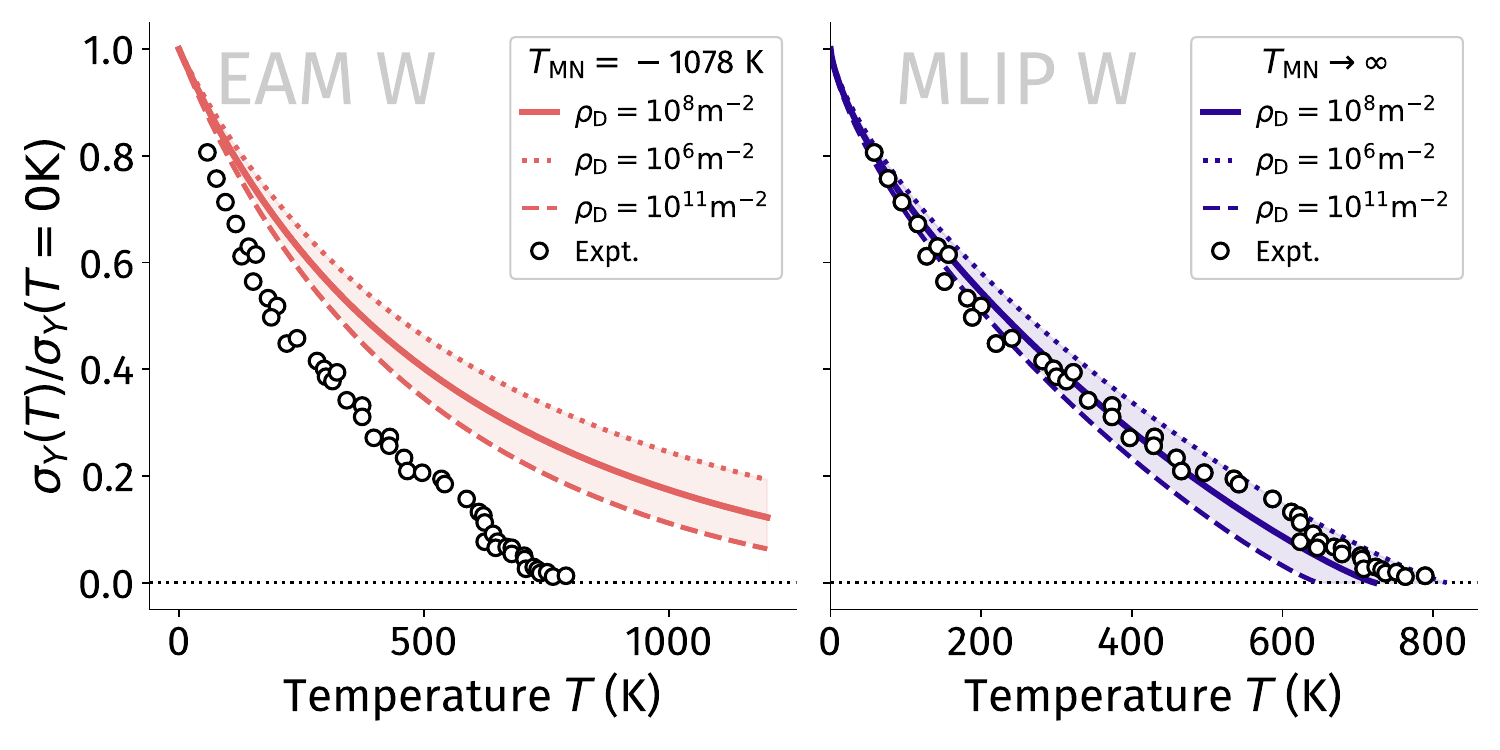}}
    \caption{\textbf{Yield stress model parameterized on atomistic calculations, compared to experimental data.} The model from Ref.~\cite{clouet2021screw} was parameterized using EAM potentials or MLIPs, in Fe and W.  Experimental data are reproduced from Refs.~\cite{keh1967plasticity,zwiesele1979temperature} for Fe and Ref.~\cite{brunner2000plastic} in W. Shaded areas indicate the variation range due to changes in dislocation density.
    }
    \label{fig:experimental_yield}
\end{figure*}

To illustrate the direct impact of entropic effects on macroscopic properties, we employed the analytical model from Ref.~\cite{clouet2021screw} to predict the thermally activated yield stress in Fe and W. The predictions are compared with experimental data from Refs.~\cite{keh1967plasticity,zwiesele1979temperature} for Fe and Ref.~\cite{brunner2000plastic} for W. The model requires no fitting parameters apart from the attempt frequency and can be fully parameterized using either MLIPs or EAM potentials (see the Methods section for details). For calculations using EAM potentials, we incorporated the inverse MN relation, $\Delta S = -\Delta H/T_{\textrm{MN}}$, with characteristic temperatures of \SI{406}{\kelvin} for Fe and \SI{1078}{\kelvin} for W as obtained in the PAFI calculations of Sec. \ref{sec:EAM}. In contrast, for calculations using MLIPs, and in agreement with the results presented in Sec.~\ref{sec:MLIP}, we assumed a constant activation entropy of $\Delta S = 6.3 \mathrm{k_B}$ for both metals. To apply the model under conditions representative of experimental settings, one parameter difficult to evaluate is the density of mobile dislocations. To address this, we considered a range of densities from $10^{6}$ to $10^{11}$ m$^{-2}$, using $10^{8}$ m$^{-2}$ as a reference value, in agreement with estimates from the literature \cite{Brunner2000,Brunner2000b,Caillard2010}.

The results are presented in Fig.~\ref{fig:experimental_yield}. As with all previous EAM and DFT calculations, the present MLIPs overestimate the Peierls stress at \SI{0}{\kelvin} compared to experiments. To eliminate this well-known discrepancy ~\cite{proville2012quantum,freitas2018quantum,proville2013prediction}, all curves were normalized by the extrapolated zero-Kelvin yield stress.
Both MLIP predictions exhibit excellent agreement with experimental data. Notably, they accurately capture the athermal temperature, above which plasticity ceases to be thermally activated. For the EAM potentials, it is interesting to observe that the prediction for Fe is rather accurate, although with deviations at both low and high temperatures, whereas in W, the EAM potential significantly underestimates the temperature dependence of the yield stress. We believe that the agreement of the EAM potential in Fe results from a compensation effect: the kink-pair nucleation enthalpy is underestimated, as seen in Fig. \ref{fig:4}, while the entropy is overestimated, as shown in Fig. \ref{fig:3}. In W, the enthalpy is better reproduced, but the entropy remains overestimated, leading to a slower decrease in the yield stress. Regardless, these results clearly highlight the ability of MLIPs to predict macroscopic yield stress without the need for fitting parameters.

\section{Discussion}
\label{sec:discussion}

The main conclusion of this work may seem unremarkable at first: MLIPs reveal that the activation entropy for dislocation glide in BCC Fe and W is constant, as assumed in the simplest models. However, this finding contrasts sharply with previous studies based on EAM potentials~\cite{warner2009origins,ryu2011entropic,saroukhani2016harnessing,esteban2020influence,gilbert2013free,zotov2022entropy,bagchi2024anomalous}, suggesting a strongly varying activation entropy indicative of pronounced anharmonic effects.
The present calculations indicate that this variability is likely an artifact of EAM potentials, which tend to produce a more rugged energy landscape compared to MLIPs and DFT. Nevertheless, we emphasize that potential energy landscape artifacts may still arise even with MLIPs~\cite{deng2025systematic}, and care must be taken to validate their reliability in each application.

Vibrational entropy arises from subtle variations in the vibrational density of states, which EAM potentials struggle to capture due to their simplified formulation. In contrast, MLIPs exhibit a consistently harmonic behavior, as demonstrated by the accurate entropy predictions across the entire temperature range when using a variational HTST approach. This stands in stark contrast to the pronounced anharmonicity inherent in EAM potentials.

While our results highlight several limitations of classical EAM potentials compared to MLIPs, we acknowledge that a formal demonstration of the origin of these limitations remains elusive. What is clear, however, is that the central-force nature of EAM potentials leads to enhanced shear softening. Additional artifacts have also been reported, such as a non-monotonic temperature dependence of elastic constants~\cite{zhong2023anharmonic} and an early departure from harmonicity~\cite{swinburne2018unsupervised}.
Although some rugged features in energy landscapes, such as those reported in recent studies on metallic glasses~\cite{zella2024ripples,li2024infinitely}, may correspond to real physical phenomena, our findings highlight the need for caution in interpreting such features, as they may also stem from artifacts due to the potential. This underscores the importance of validating potential energy landscapes against DFT data or using well-trained MLIPs, and calls for more systematic cross-comparisons between different potential formalisms in future works.

As a final note, we recall that early theories on entropic effects~\cite{zener1951theory,schoeck1965,kocks1975} proposed a connection between entropy and the temperature dependence of macroscopic properties, particularly the shear modulus $\mu$. Zener~\cite{zener1951theory}, for instance, suggested that $\Delta S \propto d\mu/dT$.
To investigate this hypothesis, we computed the temperature dependence of the shear modulus using both the EAM potential and the MLIP for Fe (see \supplementary{}). While the EAM potential predicts an anomalous increase in shear modulus with temperature—potentially aligning with an inverse compensation effect in Zener’s theory—the MLIP predicts a decreasing shear modulus, a trend inconsistent with the absence of entropic effects in this potential. These findings thus suggest that inferring an enthalpy-entropy relationship from macroscopic property variations is challenging. Instead, entropy changes likely emerge from complex changes in vibrational spectra, as previously observed in amorphous solids~\cite{gelin2020enthalpy}.

Although this study focuses on metal plasticity, the results are expected to extend to other crystalline defects, such as point defects and grain boundaries, whose kinetics may also be influenced by entropic effects.  
Caution must therefore be exercised when analyzing finite-temperature kinetics predicted from the atomic scale.

\section*{Methods}

\subsection*{Machine-learning interatomic potentials}

The machine-learning interatomic potentials are based on the Quadratic Noise Machine Learning (QNML) formalism introduced in Ref.~\cite{goryaeva2021efficient, zhong2023anharmonic, zhong2025unraveling}. This approach provides a balanced compromise between computational efficiency and predictive accuracy. 
Local atomic environments are represented by bispectrum SO(4) descriptors with maximum angular momentum $J_{\textrm{max}}=4$ and dimension $K=55$.
The cutoff distance is set to \SI{4.7}{\angstrom} for Fe and \SI{5.3}{\angstrom} for W.  
The fitting of the potentials was carried out using the \textsc{MiLaDy} package~\cite{milady}. 
We use weights in the objective loss function to control accuracy in different subsets of the database and optimize properties of interest. 
We extended existing general-purpose fitting databases for Fe and W~\cite{goryaeva2021efficient, zhong2023anharmonic} with dislocation configurations.  
For consistency, we ensured that the new configurations were computed with the same DFT parameterization as the original databases, using the VASP code.
The Mahalanobis outlier distance is used in \supplementary{} to detect out-of-distribution configurations~\cite{goryaeva2020reinforcing}.

\subsection*{Atomistic simulations of screw dislocations}

The simulation cell contains a screw dislocation of length $L = 40b$, where $b$ is the Burgers vector of the dislocation, defined as $b = \frac{1}{2}[111]$. The dislocation glides in a $\{110\}$ central plane with a glide distance $D \sim 50b$ and a cell height perpendicular to the glide plane of $H \sim 30b$. Periodic boundary conditions were applied in the glide plane to form a periodic array of dislocations~\cite{bacon2009dislocation}. The total number of atoms in the simulation cell is 96,000.  

Stress-controlled molecular dynamics simulations were performed using LAMMPS~\cite{LAMMPS} with the MLIPs. The simulations lasted from \SI{0.2}{} to \SI{0.6}{\nano\second} to ensure the computation of well-converged average dislocation velocities. The simulation time step was set to \SI{1}{\femto\second}, with a total computational cost between \SI{3e4}{} and \SI{9e4}{} CPU hours per condition.  

\subsection*{Computation of Gibbs energy barriers}

Gibbs energy barriers were calculated using the linear-scaling Projected Average Force Integrator (PAFI) method~\cite{swinburne2018unsupervised}. Transition pathways were first relaxed using the nudged elastic band (NEB) method~\cite{henkelman2000climbing} to a force tolerance of \SI{e-3}{\electronvolt\per\angstrom}. A custom parallel nudging force, designed to promote equidistant energy differences along the path, was implemented in LAMMPS~\cite{LAMMPS}.  

The PAFI approach enables the calculation of the anharmonic Gibbs activation energy at $\mathcal{O}(N)$ cost for kink-pair nucleation in bcc metals under finite applied stress along the \SI{0}{\kelvin} reaction coordinate, without requiring any a priori assumptions about finite temperature changes of the pathways. 
Anharmonic Gibbs energy calculations were performed using the PAFI package~\cite{pafi}~\cite{swinburne2018unsupervised,sato2021anharmonic}. Within this method, the configurations at a given applied stress and specific reaction coordinate are sampled at all temperatures from the same hyperplane, which is perpendicular to the initial NEB path. Furthermore, after sampling, configurations are relaxed in their hyperplanes to eliminate trajectories that might have escaped their initial potential energy valley. The computational cost ranged from \SI{5e4}{} to \SI{2.5e5}{} CPU hours per condition, mainly depending on the number of independent samplings, which typically ranges from 5 to 40 and was optimized to control numerical errors (see \supplementary{} for a discussion of error estimation in PAFI calculations).  

Hessian matrix diagonalizations were used to compute the harmonic activation entropy within the framework of the harmonic transition state theory. 
The Hessian matrices were generated using LAMMPS and subsequently diagonalized with the Phondy code to evaluate the phonon spectrum~\cite{Marinica2007,Soulie2018,Berthier2019}. The total computational cost for these calculations amounted to approximately \SI{5e4}{} CPU-hours per atomic system using the MLIP.  

\subsection*{Analytical yield stress model}

The thermally activated yield stress in Fe and W was predicted using the analytical model developed in Ref.~\cite{clouet2021screw}, which expresses the normalized yield stress as:
\begin{equation}  
\frac{\sigma_Y(T)}{\sigma_Y(T=0K)} =  \left( 1 - \left( \frac{\mathrm{k_B} T}{\Delta H_0 \left(1-\frac{T}{T_{\textrm{MN}}} \right)} \ln \left( \frac{\dot\epsilon_p}{\sqrt{\rho_D} \nu_D \lambda_P} \right) \right)^{\frac{1}{q}} \right)^{\frac{1}{p}}.
\end{equation}

This expression derives from a Kocks representation of the activation enthalpy:
\begin{equation}    
\Delta H = \Delta H_0 \left(1-\left(\frac{\sigma_Y(T)}{\sigma_Y(T=0K)}\right)^p\right)^q
\end{equation}
and a MN law for the activation entropy:
\begin{equation}    
\Delta S = \frac{\Delta H}{T_{\textrm{MN}}}.
\end{equation}

Activation enthalpies were fitted to NEB calculations using the different potentials. For Fe, we used $\Delta H_0 = \SI{0.6}{\electronvolt}$, p=0.53 and q=1.06 for the EAM potential, and $\Delta H_0 = \SI{0.8}{\electronvolt}$, p=0.87 and q=1.33 for the MLIP. For W, we used $\Delta H_0 = \SI{1.6}{\electronvolt}$, p=0.68 and q=1.02 for the EAM potential, and $\Delta H_0 = \SI{1.54}{\electronvolt}$, p=0.86 and q=1.43 for the MLIP.
The harmonic behavior predicted by the MLIPs was obtained in the limit $T_{\textrm{MN}} \rightarrow \infty$ while the inverse MN law predicted by the EAM potentials was represented with $T_{\textrm{MN}} =$ \SI{-406}{\kelvin} in Fe and \SI{-1078}{\kelvin} in W. The strain rates were set to the experimental values, $\dot \epsilon_P= \SI{5.6e-4}{\per\second}$ in Fe, and \SI{8.5e-4}{\per\second} in W. The distance between Peierls valleys is $\lambda_D = \SI{2.31}{\angstrom}$ in Fe, \SI{2.6}{\angstrom} in W. 
The attempt frequency was set to $\nu_D = \SI{6.15e13}{\per\second}$ in Fe and $\SI{1e13}{\per\second}$ in W.

\section*{Data Availability} 
The data generated in this study have been deposited in the Zenodo repository 10.5281/zenodo.15746898 \cite{zenodo}. The repository includes the Fe and W MLIPs and a minimal dataset to reproduce the results. 
Source data are provided with this paper.

\section*{Code availability}
The \textsc{MiLaDy} package is open-source software distributed under the Apache Software License (ASL) \cite{milady}. The PAFI package is available on GitHub~\cite{pafi} and is also included in the EXTRA-FIX package of the LAMMPS software.

\bibliography{bibliography}

\section*{Acknowledgments}
Emmanuel Clouet, Lisa Ventelon and Thomas Leveau are gratefully acknowledged for insightful discussions. 
AA acknowledges financial support from the Cross-Disciplinary Program on Numerical Simulation of the French Alternative Energies and Atomic Energy Commission (CEA). 
A.A., A.M.G., T.D.S and M.-C.M. acknowledge support from GENCI - (Jean-Zay/CINES/CCRT) computer centre under Grant No. A0170906973.
T.D.S gratefully acknowledges support from ANR grant ANR-23-CE46-0006-1, IDRIS allocations A0120913455, and Euratom Grant No. 633053.

\section*{Author Contributions Statement}
A.A., T.D.S., M.-C.M. and D.R. conceived the study.
B.B., M.-C.M., A.M.G. and A.A. created the training dataset and performed DFT calculations.
A.A., M.-C.M. and A.M.G. developed and fitted the machine learning potentials. 
A.A. and T.D.S. conducted the MD and PAFI simulations. 
A.A. and M.-C.M. performed harmonic entropy calculations. 
A.A and B.B. adapted the yield model. 
All authors contributed to the interpretation of the results and participated in scientific discussions throughout the study.
A.A. and D.R. drafted the initial manuscript and subsequent revisions, which were reviewed by all authors.

\section*{Competing Interests Statement}
The authors declare no competing interests.

\end{document}


\title{Supplemental Material: Activation entropy of dislocation glide \\ }

\author{Arnaud Allera}
\email[]{arnaud.allera@asnr.fr}
\affiliation{ASNR/PSN-RES/SEMIA/LSMA Centre d'\'etudes de Cadarache, F-13115 Saint Paul-lez-Durance, France}
\affiliation{Univ. Lyon, UCBL, Institut Lumi\`ere Mati\`ere, UMR CNRS 5306, F-69622  Villeurbanne, France}\affiliation{Université Paris-Saclay, CEA, Service de Recherche en Corrosion et Comportement des Matériaux, SRMP, F-91191, Gif-sur-Yvette, France}

\author{Thomas D. Swinburne}
\affiliation{Aix-Marseille Univ., CINaM, UMR CNRS 7325, Campus de Luminy, F-13288 Marseille, France}

\author{Alexandra M. Goryaeva}
\affiliation{Université Paris-Saclay, CEA, Service de Recherche en Corrosion et Comportement des Matériaux, SRMP, F-91191, Gif-sur-Yvette, France}

\author{Baptiste Bienvenu}
\affiliation{Université Paris-Saclay, CEA, Service de Recherche en Corrosion et Comportement des Matériaux, SRMP, F-91191, Gif-sur-Yvette, France}

\author{Fabienne Ribeiro}
\affiliation{IRSN/PSN-RES/SEMIA/LSMA Centre d'\'etudes de Cadarache, F-13115 Saint Paul-lez-Durance, France}

\author{Michel Perez}
\affiliation{Univ. Lyon, INSA Lyon, UCBL, MATEIS, UMR CNRS 5510, F-69621 Villeurbanne, France}

\author{Mihai-Cosmin Marinica}
\affiliation{Université Paris-Saclay, CEA, Service de Recherche en Corrosion et Comportement des Matériaux, SRMP, F-91191, Gif-sur-Yvette, France}

\author{David Rodney}
\email[]{david.rodney@univ-lyon1.fr}
\affiliation{Univ. Lyon, UCBL, Institut Lumi\`ere Mati\`ere, UMR CNRS 5306, F-69622  Villeurbanne, France}
\renewcommand\thesection{Suppl. Note \arabic{section}}

\maketitle

\onecolumngrid
\setcounter{figure}{0}
\renewcommand{\thefigure}{Supplementary Figure \arabic{figure}}
\renewcommand{\figurename}{}

\section{MLIP training}

In this work, we refitted MLIPs for Fe and W on extended databases of dislocation configurations. 
The adopted formalism is the Quadratic Noise Machine Learning (QNML) of Ref.~\onlinecite{goryaeva2021efficient}, which offers a competitive compromise between efficiency and accuracy.
All methodological details regarding the formalism and model training, descriptor parameters, and DFT parameterization can be found in the original reference.
To ensure consistency of the database, all added configurations were rescaled to the appropriate lattice parameter, and a single DFT force evaluation was performed using the same DFT parameterization as in Ref.~\onlinecite{goryaeva2021efficient}.

\subsection{Fe database}
The configurations added to the database are presented in Table~\ref{tab:dft_database} and described below:

\begin{itemize}
\item {Relaxed configurations} of dislocations and single kinks:
\begin{itemize}
    \item Relaxed easy, hard and split configurations of \textonehalf $\langle 111 \rangle$ screw dislocations, NEB Peierls barrier from Ref.~\onlinecite{bienvenu2022ab}, Hard-to-Split path from Ref.~\onlinecite{dezerald2016plastic}.
    \item Single kinks (positive $K^+$ and negative $K^-$) constructed following Ref.~\onlinecite{ventelon2009atomistic} with a total length of $6~b$.
    \item Relaxed configuration and migration barrier of $\langle 100 \rangle$ dislocations following the methodology of Refs.~\onlinecite{bienvenu:tel-03917543,bienvenu2022ab2}. 
\end{itemize}

\item {Finite temperature configurations} to improve the description of the potential energy landscape around energy minima and transition states. MD trajectories were generated using the linear MLIP of Ref.~\onlinecite{goryaeva2021efficient}.
\begin{itemize}
    \item Snapshots of MD trajectories along a constant-pressure heating ramp from \SI{100}{K} to \SI{800}{K} at \SI{2.3}{\kelvin\per\pico\second}, starting from the easy \textonehalf $\langle 111 \rangle$ screw dislocation configuration and single-kink configurations listed above.
    \item Thermalized configurations obtained using the PAFI method (i.e. constrained MD normal to the reaction pathway) along the Peierls barrier at \SI{100}{\kelvin} and \SI{300}{\kelvin}. This method allows to efficiently sample configurations that are evenly distributed along the finite temperature migration path.
\end{itemize}
\end{itemize}

Potential fitting was performed using the \textsc{MiLaDy} package \cite{milady}.
Per-class weights were adjusted to reflect the importance of the different classes in the database, in order to optimize materials properties. 
The obtained potential is characterized by a mean absolute error on forces of \SI{2.5e-2}{\electronvolt\per\angstrom} and a mean absolute error on the energy per atom of \SI{7.9e-3}{\electronvolt}, on par with other MLIPs.

\begin{table}[h!]
\begin{tabular}{clcccc}
\hline
DB class & \multicolumn{1}{c}{Contents} & \begin{tabular}[c]{@{}c@{}}Atoms \\ per cell\end{tabular} & \begin{tabular}[c]{@{}c@{}}Fitted \\ properties\end{tabular}  & \begin{tabular}[c]{@{}c@{}}$n_e$ + $n_f$ + $n_S$ \\ train/test\end{tabular} & \begin{tabular}[c]{@{}c@{}}Configurations \\ train/test\end{tabular} \\ \hline
1 & \begin{tabular}[c]{@{}l@{}}Relaxed 135 atoms \textonehalf$\langle 111 \rangle$ SD \\ (easy, hard, split, Peierls barrier)\end{tabular} & 135 & E F & 3248/0 & 8/0 \\
2 & Relaxed Hard-Split path \textonehalf$\langle 111 \rangle$ SD & 135 & E F S & 1648/0 & 4/0 \\
3 & Easy \textonehalf$\langle 111 \rangle$ screw NPT \SI{100}{\kelvin}-\SI{800}{\kelvin} & 135 & F S & 2055/411 & 5/1 \\
4 & Peierls barrier \textonehalf$\langle 111 \rangle$ SD PAFI 100K & 135 & E F S & 1236/412 & 3/1 \\
5 & Peierls barrier \textonehalf$\langle 111 \rangle$ SD PAFI 300K & 135 & F S & 1233/411 & 3/1 \\
6 & relaxed 6b \textonehalf$\langle 111 \rangle$ SD K+ & 720 & E F S & 2167/0 & 1/0 \\
7 & MD NPT 6b \textonehalf$\langle 111 \rangle$ SD K+ \SI{100}{\kelvin}-\SI{800}{\kelvin} & 720 & E F S & 10835/2167 & 5/1 \\
8 & Relaxed 6b \textonehalf$\langle 111 \rangle$ SD K- & 765 & E F S & 2302/0 & 1/1 \\
9 & MD NPT 6b \textonehalf$\langle 111 \rangle$ SD K- \SI{100}{\kelvin}-\SI{800}{\kelvin} & 765 & E F S & 11510/2302 & 5/1 \\
10 & Relaxed and NEB migration $\langle 100 \rangle$ dislocation & 135 & E F S & 3642/0 & 6/0  \\ \hline
 & \textbf{Total} & & & \textbf{39876}/5703 & 41/5 \\
 \hline
\end{tabular}

\caption{\textbf{Screw dislocation (SD) configurations added to the Fe database of Ref.~\onlinecite{goryaeva2021efficient}. }
Each database (DB) class corresponds to a category of atomic systems.
Depending on the category, the energy (E), force (F) and/or stress (S) was used or ignored for fitting.
The total number of energy ($n_e$), force ($n_f$), and stress ($n_s$) data points for each class is reported, and their distribution between training and testing sets is indicated.
The corresponding number of configurations (i.e. simulation supercells) is also indicated for each class.
}
\label{tab:dft_database}
\end{table}

\subsection{W database}

In W, we used a more limited amount of dislocation configurations, targeted to tune the properties of the potential at minimum cost.
The original W database of Ref.~\cite{goryaeva2021efficient} was already extended in Ref.~\cite{zhong2024machine} for better performance in free energy calculations of defects.
Building on this work, the present additions to the database are focused on screw dislocations, and consist of two parts:
\begin{itemize}
    \item Peierls barrier for \textonehalf$\langle 111 \rangle$ screw dislocation, relaxed using the original QNML potential \cite{goryaeva2021efficient}
    \item  Thermalized configurations sampled along the Peierls barrier, using the PAFI method at \SI{300}{\kelvin}, \SI{600}{\kelvin}, \SI{900}{\kelvin}. This procedure allows to sample states near minimum-free-energy pathways. We applied standard CUR decomposition techniques \cite{mahoney2009cur} on the atomic descriptors of the sampled configurations to select a set of configurations that promote diversity in atomic environments.  
\end{itemize}
The resulting QNML potential accurately reproduces the migration trajectory of the dislocation core, height of Peierls barrier and variation of the core eigenstrains as computed by DFT~\cite{bienvenu2022ab}, as demonstrated in section \ref{supp:potentials-properties}.

\begin{table}[h!]
\begin{tabular}{clcccc}
\hline
DB class & \multicolumn{1}{c}{Contents} & \begin{tabular}[c]{@{}c@{}}Atoms \\ per cell\end{tabular} & \begin{tabular}[c]{@{}c@{}}Fitted \\ properties\end{tabular}  & \begin{tabular}[c]{@{}c@{}}$n_e$ + $n_f$ + $n_S$ \\ train/test\end{tabular} & \begin{tabular}[c]{@{}c@{}}Configurations \\ train/test\end{tabular} \\ \hline
1 & \begin{tabular}[c]{@{}l@{}}Relaxed 135 atoms \textonehalf$\langle 111 \rangle$ SD \\ (easy, Peierls barrier)\end{tabular} & 135 & E F S & 5768/0 & 14/0 \\
2 & Peierls barrier \textonehalf$\langle 111 \rangle$ PAFI 300K, 600K, 900K & 200 & E F S & 6180/0 & 15/0  \\ \hline
 & \textbf{Total} & & & \textbf{11948}/0 & 29/0 \\
 \hline
\end{tabular}
\caption{\textbf{Screw dislocation (SD) configurations added to the W database of Ref.~\onlinecite{goryaeva2021efficient}.}
}
\label{tab:dft_database_W}
\end{table}
\newpage

\FloatBarrier
\section{Thermo-elastic properties of interatomic potentials}

The lattice parameter and elastic moduli directly influence the mechanical properties of metals.
In the framework of continuum models, an appropriate treatment of their evolution with temperature is classically considered essential for a realistic prediction of dislocation glide kinetics~\cite{kocks1975,schoeck1965}.
We report the temperature evolution of the lattice parameter $a(T)$ and shear modulus $\mu(T)$, computed with both the EAM and MLIP for Fe used in this study.

Simulations presented in this section are performed in a cell oriented along the same directions as the MD simulations discussed in the main text, i.e. $\textbf{x} = [111]$, $\textbf{y} = [\bar 1 2 \bar 1]$, $\textbf{z} = [\bar 1 0 1]$, and containing 6912 and 1728 bcc iron atoms, for the EAM and MLIP respectively.
Molecular dynamics simulations consist in two stages for each temperature, during which a shear stress $\tau_{xz}$ is imposed: a ramp of N steps in the NPT ensemble, to reach a target temperature, followed by N more NPT steps where the temperature is held constant and where the cell deformation is averaged, with N=\SI{5e4}{} steps and a \SI{0.5}{\femto\second} timestep. This process is repeated iteratively for all temperatures. 

\begin{figure}
    \centering
    \includegraphics[width=0.6\textwidth]{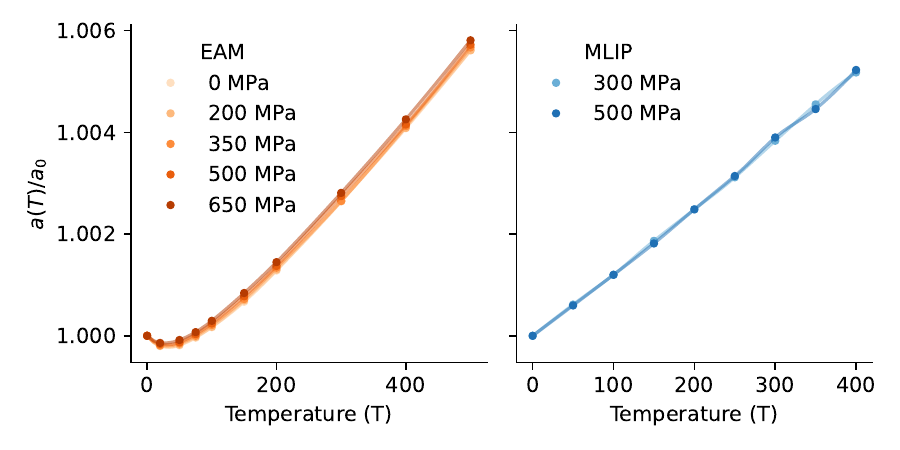}   
    \caption{\textbf{Temperature evolution of the lattice parameter $a(T)$ predicted in Fe with an EAM potential (left) and a MLIP (right).}}
    \label{fig:lattice_cst}
\end{figure}

The obtained lattice parameter is shown in \ref{fig:lattice_cst}.
With the EAM potential, the lattice parameter decreases at low temperature, before increasing above ~\SI{50}{\kelvin}. This low-temperature negative thermal expansion coefficients is an artifact of EAM potentials based on the Mendelev potential~\cite{mendelev2003development}, as reported in Ref.~\cite{chiesa2009free}.
Conversely, the MLIP predicts a lattice parameter monotonically increasing with the temperature, in excellent agreement with experimental data, similar to the originally published version of the MLIP~\cite{goryaeva2021efficient}. 
We verified that the lattice parameter is not affected by the applied shear stress $\tau_{xz}$ by testing an array of values of $\tau_{xz}$ in the domain investigated in our study, shown in \ref{fig:lattice_cst}.

\begin{figure}
    \centering
    \includegraphics[width=0.6\textwidth]{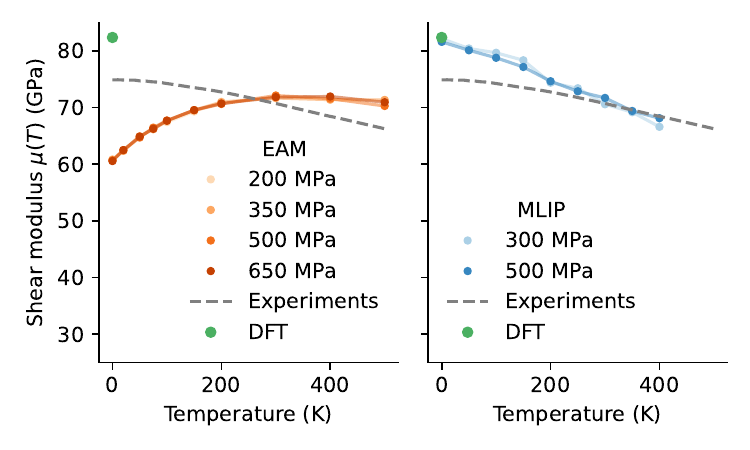}   
    \caption{\textbf{Temperature evolution of the shear modulus $\mu(T)$ predicted in Fe with an EAM potential (left) and a MLIP (right).} Comparison is made with DFT~\cite{goryaeva2021efficient} and experimental ~\cite{adams2006elastic} data.}
    \label{fig:bulk_mod}
\end{figure}

Given the orientation of our simulation cell, the screw dislocation is driven by a shear stress in the $xz$ direction. The associated shear modulus is $\mu = \frac{1}{3} (C_{11} -C_{12} + C_{44}) = \frac{1}{3}(2C'+C_{44})$, with $C' = \frac{1}{2} (C_{11}-C_{12})$.
The shear modulus can be computed in MD simulations under an applied $\tau_{xz}$ by measuring $\epsilon_{xz}$, as $\mu = \tau_{xz}/\epsilon_{xz}$.
Results obtained with both interatomic potentials are shown in \ref{fig:bulk_mod}, along with experimental data from \cite{adams2006elastic} and DFT calculations of \cite{goryaeva2021efficient}.
Again, the EAM potential shows a qualitative disagreement with experiments, with a shear modulus which increases with temperature, until $\approx$\SI{300}{\kelvin}, where it becomes constant. 
For the MLIP, we find a monotonic decrease of $\mu$, in qualitative agreement with experiments, matching DFT at \SI{0}{\kelvin}, although the rate of decrease is faster than in the experiments.
We checked that the computed shear modulus is not affected by the value of $\tau_{xz}$, for both potentials.
Note that the slight roughness of the MLIP curves is due to our choice of a smaller system compared to the EAM to limit the computational cost, resulting in a larger variance.

\FloatBarrier
\section{Error estimation in PAFI calculations}
\label{app:error-pafi}
The PAFI approach \cite{swinburne2018unsupervised} collates an ensemble of $N_E$ independent estimations,  $g_i(z)$, $i\in[1,N_E]$ of the free energy gradient $\partial_z G_(z, \tau, T)$. As each sample duration $\tau_S=N_S\tau_c$ is much longer than the phonon decorrelation time $\tau_c$, 
the error of each individual sample will scale approximately as $\sigma/\sqrt{N_S}$. 
As a result, the ensemble average $\bar{g}(z) = \sum_i g_i(z)/N_E$ will estimate $\partial_z G(z, \tau, T)$ with an error $\bar{\sigma}=\sigma/\sqrt{N_SN_E}$. 
It is simple to show that the ensemble variance $\sigma^2_E=\sum_i g^2_i(z)/N_E-\bar{g}^2(z)\to\sigma^2/\sqrt{N_S}$ as $N_E\to\infty$, 
meaning the error on the free energy gradient is given by $\bar{\sigma} = \sigma_E/\sqrt{N_E}$. This gradient error is integrated to give the error on the barriers presented in the main text. 
A more detailed discussion with derivation and examples can be found on the PAFI code repository \cite{pafi}.

\FloatBarrier
\section{Anharmonicity of Gibbs activation energy }

In this section, we provide details on the temperature dependence of the Gibbs activation energy shown in Figs.~3 and 4 of the main text. The calculations presented here were performed using the EAM and MLIP potentials for Fe.

\subsection{EAM potential}

\begin{figure}[h!]
    \centering
    \includegraphics[width=0.90\linewidth]{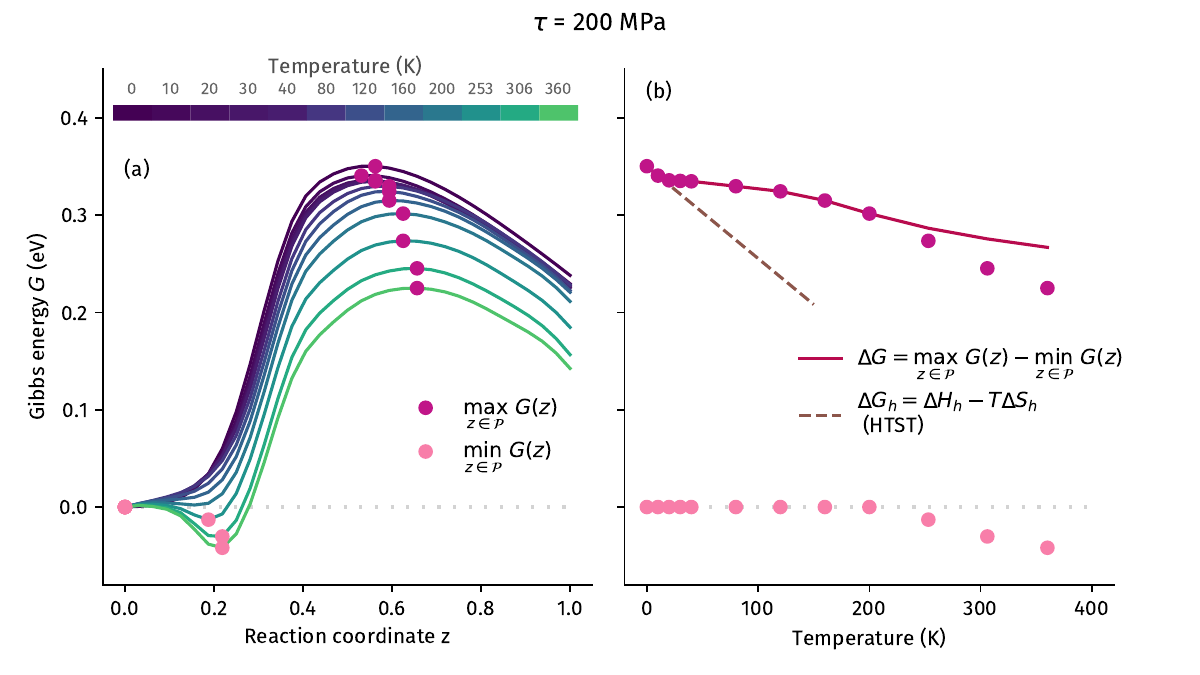}
    \caption{\textbf{Gibbs energy calculated with the PAFI method and the EAM potential for Fe.}
    (a) Gibbs energy profiles at \SI{200}{\mega\pascal} and temperatures between 0 and 360K.
    Energy minima and maxima are marked by pink and purple circles, respectively.
    (b) Temperature evolution of the minimum (pink circles), maximum (purple circles) Gibbs energy and corresponding energy barrier $\Delta G$. 
    For reference, the harmonic approximation is shown as a dashed line.}
    \label{fig:supp2.2}
\end{figure}

\ref{fig:supp2.2} shows Gibbs energy profiles obtained with the EAM potential at different temperatures and \SI{200}{\mega\pascal}. In \ref{fig:supp2.2}~(a), the horizontal axis shows the reaction coordinate $z$ along the transition path $\mathcal{P}$. 
We note $z$ the normalized distance from the initial configuration to the configurations along the zero-Kelvin NEB path, from which hyperplanes are defined to perform the sampling at finite temperature~(see Ref.~\onlinecite{pafi,swinburne2018unsupervised} for details).

As temperature increases, the Gibbs energy barrier decreases, and the reaction coordinates of the minimum and maximum energy states shift towards larger values (pink and purple circles). 
The same energies are plotted in Fig~\ref{fig:supp2.2}~(b) as a function of temperature (circles). 
We also added the Gibbs activation energy $\Delta G$ (full line), which is the difference between the maximum and minimum Gibbs energies.

Below \SI{200}{\kelvin}, the minimum-Gibbs energy configuration is the initial configuration at $z=0$. 
This configuration is used as a reference to compute the Gibbs energies, thus its Gibbs energy is zero at all temperatures by construction. 
Above \SI{200}{\kelvin}, the minimum-energy configuration is no longer the initial configuration. It gradually shifts to higher reaction coordinates, with a Gibbs energy that is now below 0. 
Meanwhile, the maximum Gibbs energy decreases highly non-linearly with temperature and the position of the activated state also varies with temperature. 

In the conventional (non-variational) HTST, $\Delta G_h$ is the difference of harmonic Gibbs energies between the initial and activated states of the zero-Kelvin MEP. 
The corresponding prediction is shown as a dashed line in Fig~\ref{fig:supp2.2}~(b), where we recover the marked deviation from $\Delta G$ starting at very low temperatures (above 20 K) discussed in the main text. 
Since the dislocation merely fluctuates around its straight configuration at $z$=0, we expect that the anharmonic decrease of $\Delta G$ below 200 K with the EAM potential is mainly due to an evolution of the activated state, corresponding to a widening of the activated kink-pair reported in Ref.~\onlinecite{swinburne2018unsupervised}.

\subsection{MLIP}
\label{supp:vhtst_mlip}

\begin{figure}[h]
    \centering
    \begin{minipage}{0.8\linewidth}
        \includegraphics[width=\linewidth]{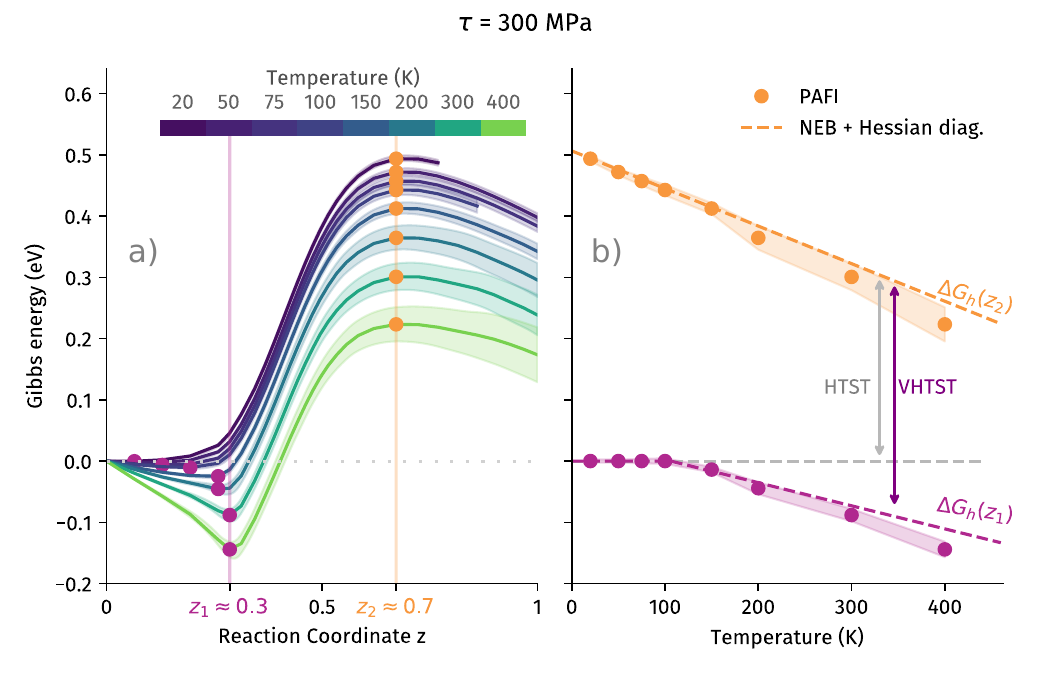}
        
    \end{minipage}
    \begin{minipage}{0.8\linewidth}
        \includegraphics[width=\linewidth]{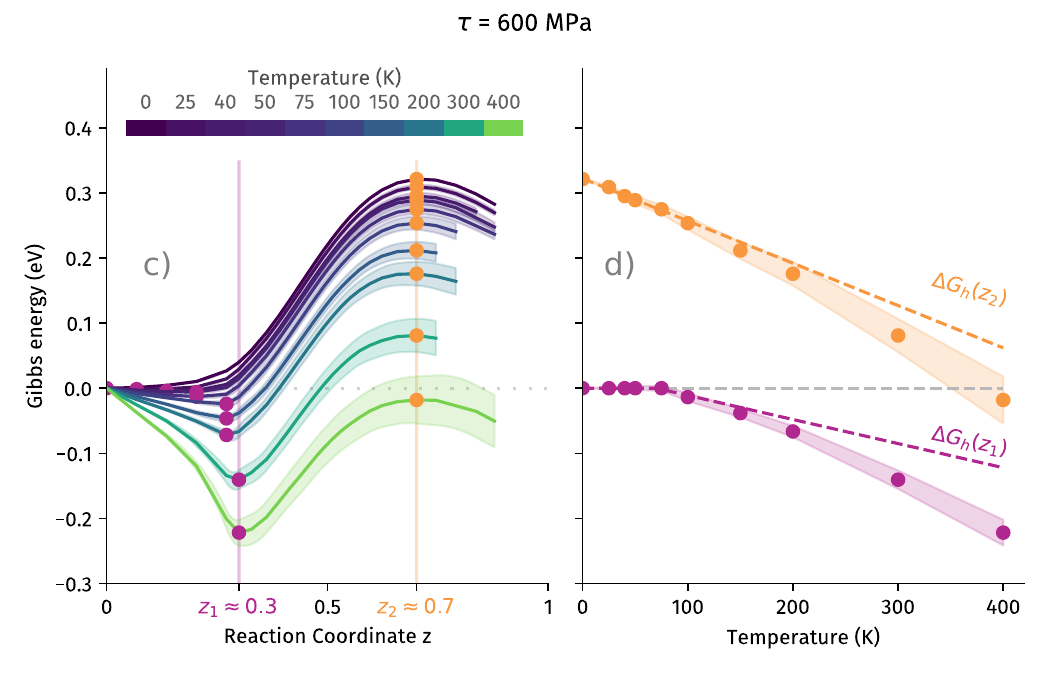}
        
    \end{minipage}

    \caption{\textbf{Gibbs energy calculated with the PAFI method and the MLIP for Fe.}
    (a) Gibbs energy profiles at temperatures between 0 and 400K. (b) Temperature evolution of the Gibbs energy $G(z_{\{1,2\}}, \tau, T)$,  computed at the position of the minimum ($z=0.3$) and maximum ($z=0.7$) Gibbs energies at high temperature.
    Calculations were performed at \SI{300}{\mega\pascal} (top row) and \SI{600}{\mega\pascal} (bottom row).
    Dashed lines on the right panel show the harmonic approximation.
    Shaded regions represent the error estimation of the PAFI method, as described in \ref{app:error-pafi}.
    }
    \label{fig:supp2}
\end{figure}

The same analysis is performed with the MLIP for Fe in \ref{fig:supp2}. 
From the Gibbs energy profiles (left column), we see that the coordinate of the energy minima shifts from $z=0$ to approximately $z_1=0.3$ at temperatures above \SI{100}{\kelvin}, while the energy maxima remain close to $z_2=0.7$. 

Plotted as a function of temperature (right column), the minimum Gibbs energy thus remains 0 as long as the minimum-Gibbs energy configuration is the initial configuration, i.e. below 100 K, and then decreases below zero, as with the EAM potential. 
For the maximum energy, we represent the harmonic prediction of the Gibbs energy at $z=0.7$ in orange (computed from the enthalpy and harmonic entropy of the zero-Kelvin NEB configuration at $z=0.7$). 
This prediction fits the PAFI calculations particularly well below about \SI{100}{\kelvin}, and fairly well at all temperatures. 
As reported in the main text, the HTST is thus valid up to about \SI{100}{\kelvin}, i.e. as long as the minimum energy configuration remains at $z=0$ and the maximum energy decreases linearly with temperature. 
Above \SI{100}{\kelvin}, the minimum-energy configuration shifts to $z_1=0.3$ and its energy decreases below 0, leading to a breakdown of the HTST discussed in the main text.

However, to extend the harmonic regime, one can use the variational HTST (VHTST)~\cite{truhlar1984variational}, where the energy barrier is computed  between the two configurations that maximize the energy barrier at every temperature. 
Within the VHTST, the Gibbs energy difference is thus computed in the harmonic approximation between the configurations at $z=0$ and $z=0.7$ below about \SI{100}{\kelvin}, and between $z=0.3$ and $z=0.7$ above this temperature. 
In \ref{fig:supp2} (right column), we plotted as a pink dashed line the harmonic prediction of the Gibbs energy of the configuration at $z=0.3$ (computed again from the enthalpy and harmonic entropy of the zero-Kelvin NEB configuration, at $z=0.3$). 
The agreement with the PAFI calculations of the minimum energy is very satisfying up to about 200 to 300 K, above which the non-linear decrease of the energy becomes significant.

In conclusion, the MLIP predicts a quadratic potential energy landscape around the transition path, both at zero and finite temperatures, allowing for an accurate prediction of the VHTST up to high temperatures. 
We have seen that the main effect explaining the discrepancy between PAFI calculations and HTST is a shift of the minimum-Gibbs energy configuration, which is no longer the initial configuration above 100 K. 
We have checked that at $z=0.3$, the kink pair has not yet formed on the dislocation. 
The latter thus fluctuates around an average straight configuration higher in the Peierls valley than the initial configuration. At this position, we can expect that the dislocation visits softer regions of the potential energy landscape, associated with a larger vibrational entropy than the initial configuration. This increase in $\Delta S(z)$ starts before the activated kink-pair forms, i.e. before $\Delta H(z)$ increases significantly. As a reference at \SI{300}{\mega\pascal}, $\Delta S(z=0.3) = 4.7$ k$_B$, which represents $70\%$ of the saddle entropy, $\Delta S (z=0.7) = 6.3$ k$_B$, while $\Delta H(z=0.3)=0.05$ eV, only about $10\%$ of the enthalpy barrier. This delay between entropy and enthalpy rise implies at high temperatures the existence of configurations with a negative Gibbs energy, which occurs with the MLIP at around 100 K. The same effect exists with the EAM potential, but at higher temperatures (200 K).

As a conclusion, the origin of anharmonicity with the MLIP is the  vibrational entropy increasing faster than the enthalpy 
along the transition path near the initial state, 
resulting in configurations of negative Gibbs energy,   
while with the EAM potential, the main effect is a widening of the activated state. 

\subsection{Thermal expansion}

\begin{figure}[h]
    \centering
    \includegraphics[width=.5\linewidth]{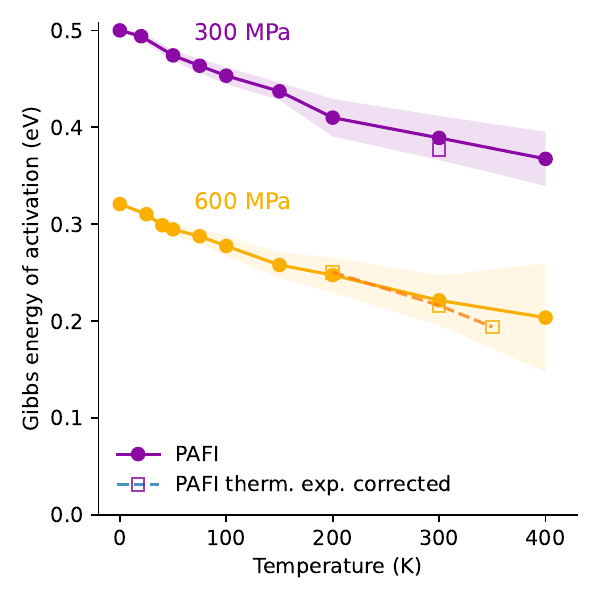}
    \caption{\textbf{Effect of thermal expansion on Gibbs activation barriers.} PAFI calculations performed in Fe in non-rescaled (full symbols) and rescaled (open symbols) cells to account for thermal expansion. Shaded regions represent the error estimation of the PAFI method, as described in \ref{app:error-pafi}.}
    \label{fig:supp3}
\end{figure}

The PAFI calculations of the main text were performed in constant-volume cells corresponding to the zero-Kelvin lattice parameter of the potentials, $a_0$. 
In order to check a potential effect of lattice expansion, we performed a few additional PAFI calculations in cells rescaled according to the law $a_0(T) = a_0(\SI{0}{\kelvin})(1+\SI{1.093e-5}{\per\kelvin}T+\SI{5.117e-9}{\per\square\kelvin}T^2)$, which was adjusted for the present MLIP following the method of Ref.~\onlinecite{zhong2023anharmonic}. 
The result is shown in \ref{fig:supp3}, where only a slight change of the Gibbs activation energy is observed. 
This is at odds with the case of cross-slip in FCC metals considered in Ref. \onlinecite{wang2023stress}, where strong thermal expansion effects were reported.

\section{PAFI calculations in W}

In order to assess potential- or material-specific effects rather than a general trend of EAM potentials, we performed  calculations in W, using a classical EAM potential denoted \texttt{eam4} in Ref.~\cite{marinica2013interatomic}, and a W MLIP refined for improved dislocation properties (see \ref{supp:potentials-properties}).

\subsection{EAM potential}
\begin{figure}[htbp]
    \centering
    \includegraphics[width=0.5\linewidth]{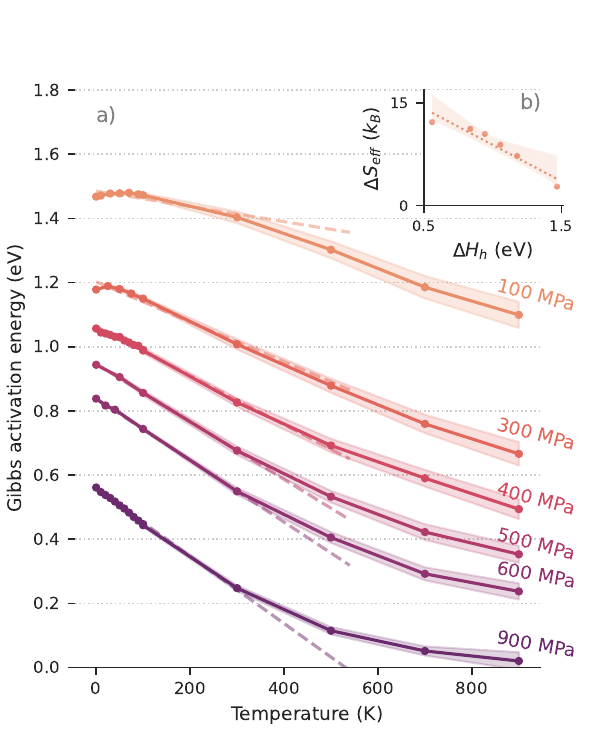}
    \caption{\textbf{Gibbs activation energy of kink-pair nucleation in W modeled by an EAM potential}. (a) PAFI calculations (symbols) at different temperatures and applied stress. The dashed lines indicate least-squares fits to linear models, below \SI{300}{\kelvin}. (b) Effective entropy determined from slopes of linear regressions at different levels of stress, as a function of the activation enthalpy. The dotted line indicates the linear fit $\Delta S_{eff}= -\Delta H_h/\SI{1078}{\kelvin}+19\mathrm{k_B}$. 
    Shaded regions in (a) represent the error estimation of the PAFI method, as described in \ref{app:error-pafi}.
    Shaded regions in (b) represent a confidence interval for linear fit, obtained by standard statistical resampling.}
    \label{fig:PAFI-EAM-W}
\end{figure}

The simulation setting and conditions are similar to Fe, here using 120 independent samplings per condition.
The results, presented in \ref{fig:PAFI-EAM-W} a), are fully consistent with the calculations in Fe discussed in the main text.
The Gibbs activation energy is again varying non-linearly in the investigated temperature-stress domain, which goes up to \SI{900}{\kelvin}.
Below \SI{300}{\kelvin}, a linear regime is observed, where a slope $\Delta S_{eff}$, representing the effective activation entropy, is extracted and plotted in \ref{fig:PAFI-EAM-W} b), 
showing an inverse Meyer-Neldel trend with a characteristic temperature $T_{MN}=-\SI{1078}{\kelvin}$, i.e. approximately $T_{MN}\approx -0.3\times T_m$, where $T_m$ is the melting temperature of W. 
This ratio is close to the $T_{MN}\approx-0.2\times T_m$ found in iron, especially considering the large uncertainty on determining these slopes.
In the low-temperature limit, nonlinearities are seen, indicating anharmonicity starting at temperatures as low as about \SI{20}{\kelvin}. 
This observation is consistent with both our calculations in Fe and previous calculations in W using the same EAM \cite{swinburne2018unsupervised}.
Similar to the Fe EAM potential investigated in this work, the present W EAM potential shows incorrect thermo-elastic properties, observed as a common feature in empirical potentials of bcc W in Ref.~\cite{zhong2023anharmonic}.

\subsection{MLIP}
We performed similar PAFI calculations using the newly developed MLIP for W.
Due to the higher computational cost of this potential, the number of workers per run ranged from 10 at low temperatures to 20 at \SI{900}{\kelvin}, resulting in a total computational cost of \SI{5e4}{} to \SI{e5}{} CPU hours per condition.

PAFI calculations were conducted at two stress levels, \SI{300}{\mega\pascal} and \SI{600}{\mega\pascal}, over a temperature range from zero to \SI{900}{\kelvin}.
The Gibbs activation energies obtained exhibit a behavior very similar to that observed for Fe:
$\Delta G$ decreases linearly up to a transition temperature in the range of \SI{100}{\kelvin}-\SI{300}{\kelvin}, where the slope changes, after which the linear trend continues up to the highest temperature investigated, \SI{900}{\kelvin}.
Notably, the entropy shows only a weak dependence on applied stress, reinforcing our key finding of stress-independent entropy.

The systematic comparison performed in this section, using a consistent computational procedure, provides valuable insights and establishes a trend across bcc metals.

\begin{figure}
    \centering
    \includegraphics[width=0.5\linewidth]{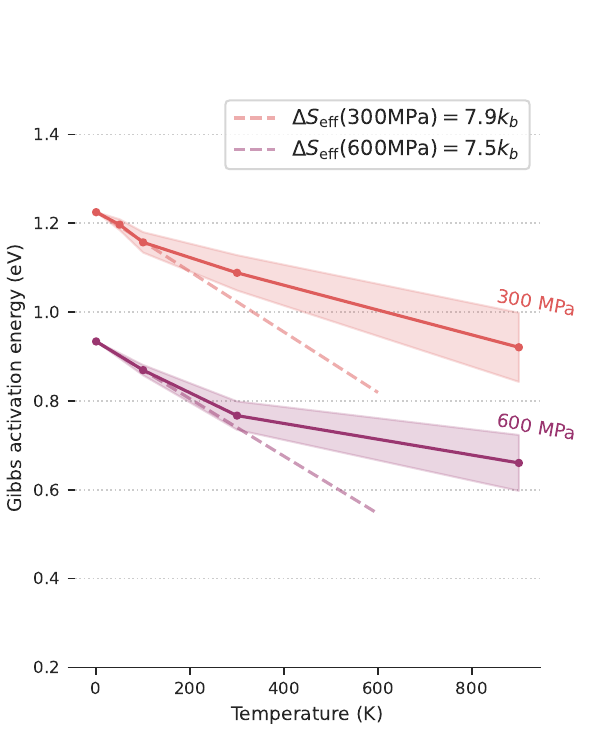}
    \caption{\textbf{Gibbs activation energy of kink-pair nucleation in W modeled by a MLIP} PAFI calculations (symbols) at different temperatures and applied stress. The dashed lines indicate least-squares fits to linear models, below \SI{200}{\kelvin}. 
    Shaded regions represent the error estimation of the PAFI method, as described in \ref{app:error-pafi}.}
    \label{fig:enter-label}
\end{figure}

\section{Out-of-distribution detection in large scale simulations}
\label{supp:OOD}

\paragraph{Methods}

Assessing the reliability of surrogate models, such as the MLPI employed in this study, is crucial for establishing confidence in simulation results.  
A fundamental metric in this regard is uncertainty quantification, which evaluates whether a given prediction during simulation  
falls within the interpolation domain of the training data or constitutes an extrapolation.  
This extrapolation challenge is particularly pronounced in the large-scale MD simulations presented here. The extensive atomic systems and long-time trajectories used to sample the free energy landscape of dislocation migration introduce significant statistical diversity. Consequently, local atomic configurations may emerge that deviate substantially from those  
represented in the training dataset.

A variety of techniques exists for estimating the degree of extrapolation, alongside providing a quantitative measure of  
prediction uncertainty. These methods range from standard textbook Bayesian models \cite{Book_Rasmussen, deringer2019machine, deringer2021gaussian}  
to recent advancements in the context of machine learning force fields \cite{xie_flare_2021, vandermause_flare_2020, swinburne2024misspecification, detlefsen_reliable_2019, neal_modern_2019, kellner_uncertainty_2024}.  
Assessing extrapolation can be formulated as an outlier detection problem, which is particularly well-suited for the ML force fields  
employed in this study. Unlike traditional interatomic potentials, these models are linear models built on a descriptor representation of atomic environments, facilitating efficient statistical analysis in high-dimensional feature space. Here, we adopt the methodology outlined in Refs.~\cite{goryaeva2020reinforcing, swinburne2023coarse}, which uses statistical distance metrics to quantify deviations from the training data distribution.

In this framework, the reliability of an ML force field for a given application can be gauged by a 1D or multi-dimensional statistical distance with respect to some training database.
In general, the dimensionality of the statistical distance corresponds to the modality of the training database- a unimodal
dataset could be described by a 1D distance, whilst in the 
general case some clustering algorithm should identify the dominant modes and the multi-dimensional statistical distance is a distance from each mode. The distance from each mode is normalized by the modes' covariance.

Here, the data is partitioned employing the  Gaussian mixture model (GMM) \cite{Jordan1994, Marin2011} on the training data. 
The data will be denoted as $\mathbf{x} \in \mathbb{R}^D$ and is written as a mixture of $n_g$ Gaussian distributions of dimension $D$, mean $\mathbf{m}_i$ and covariance matrix $\mathbf{S}_i$:
\begin{equation}
p(\mathbf{x}) = \sum_{i=1,n_g} p(\mathbf{x} |\mathbf{m}_i, \mathbf{S}_i) p_i ,
\end{equation}
where  $p_i$ is the mixture weight for component $i$.  
The optimal set of parameters for the mean $\mathbf{m}_i \in \mathbb{R}^D$, the covariance $\mathbf{S}_i \in \mathbb{R}^{D\times D}$, $p_i$ can be obtained using standard maximum likelihood optimization.
Local atomic environments $\mathbf{x}$ present in the training database are represented by BSO4 descriptors \cite{bartok2013representing} ($j_{max}=4$, $r_{cut}=\SI{4.7}{\angstrom}$). 
The distribution of descriptors associated with the database is modeled by two Gaussian distributions $G_1$, $G_2$, fitted by a 2-class Gaussian mixture model. To each Gaussian is associated a Mahalanobis statistical distance, characterizing the distance in descriptor space between $G_i$ and a sample $\mathbf{x}$~\cite{goryaeva2020reinforcing}:

\begin{equation}
    d_{maha}(G_i, \mathbf{x}) = \sqrt{({\mathbf{x}}-\mathbf{m}_i)\mathbf{S}_i^{-1}({\mathbf{x}}-\mathbf{m}_i)^\top} 
\end{equation}
This dual-sided localization of the data enables rapid positioning within the descriptor space. Overlap of the database local atomic environments with simulated data ensures reliable predictions. 
An absence of overlap does not necessarily exclude realistic results, but it also does not ensure them.

\begin{figure}[h]
    \centering
    \includegraphics[width=0.99\textwidth]{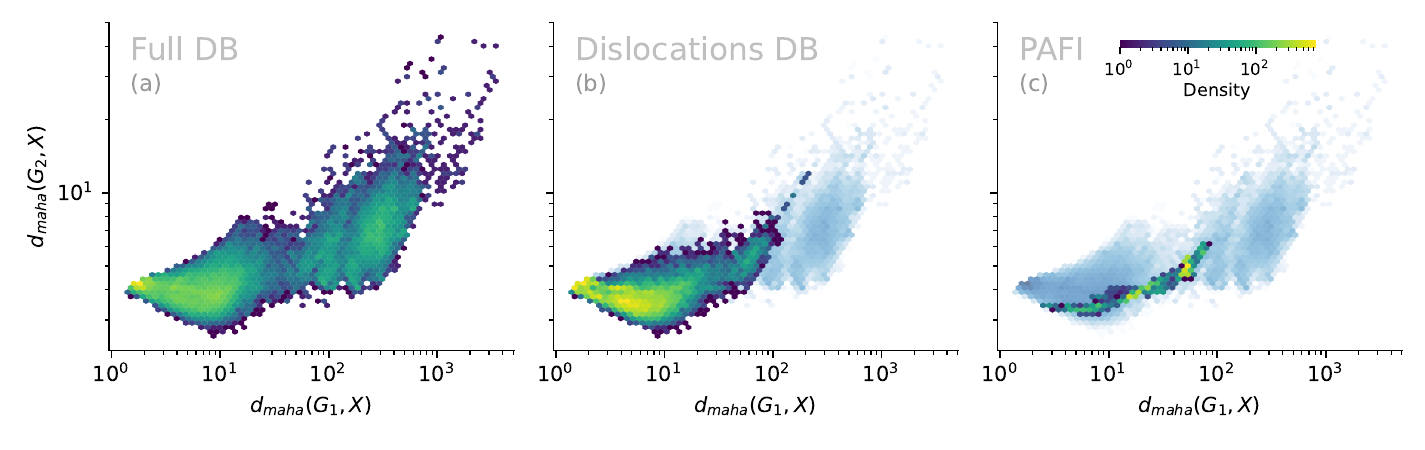}
    \caption{\textbf{Heatmaps of statistical distances from the MLIP training database, for different sets of atomic configurations.} Distributions of statistical distances, computed for (a) the full training database, (b) dislocation configurations from Table~\ref{tab:dft_database}, and (c) ensemble-averaged finite temperature dislocation configurations discovered in large-scale PAFI calculations, presented in Fig.~2.~(d).
    The color scale represents the relative density of points in logarithmic scale, from blue (lowest) to yellow (highest). 
    A global color scale is used in all panels. 
    On panels (b) and (c), heatmaps are superimposed on the full training database distances distribution (in light blue) for easier comparison.
    }
    \label{fig:database_maha}
\end{figure}

\paragraph{Results}
The distribution of statistical distances in the original database from \cite{goryaeva2021efficient}, shown in \ref{fig:database_maha}~(a) for the Fe MLIP, covers a wide range of atomic configurations, mostly contiguous in descriptor space.
To highlight the domain relevant in dislocation simulations, we present the distribution of statistical distances in the database extension proposed in this work (described in Table~\ref{tab:dft_database}) in \ref{fig:database_maha}~(b).
The database extension increases the density of data points within the domain already covered in the dataset of Ref.~\cite{goryaeva2021efficient},  
thereby improving the reliability of predictions for dislocation configuration properties.  
This reinforcement enhances the model’s interpolation capabilities, reducing uncertainty and increasing confidence in the computed results.

In order to locate the domain sampled in Gibbs energy calculations by the PAFI method and detect out-of-distribution samples, we represent the statistical distances to the training database of all atomic configurations extracted from the PAFI calculations based on the Fe MLIP, in \ref{fig:database_maha}~(c).
Atomic configurations are averaged over the ensemble of configurations sampled along each MD trajectory.
In our large-scale PAFI calculations, atoms near the dislocation core experience the largest displacements, while other atoms move only marginally and have nearly equivalent local environments.
To avoid data redundancy, we limit the analysis to atoms located within a cylinder of radius \SI{5}{\angstrom}, periodic in the [111] direction, centered on the initial position of the dislocation line, i.e. enclosing the 720 atoms closest to the dislocation core.

\begin{figure}
    \centering
    \includegraphics[width=0.9\textwidth]{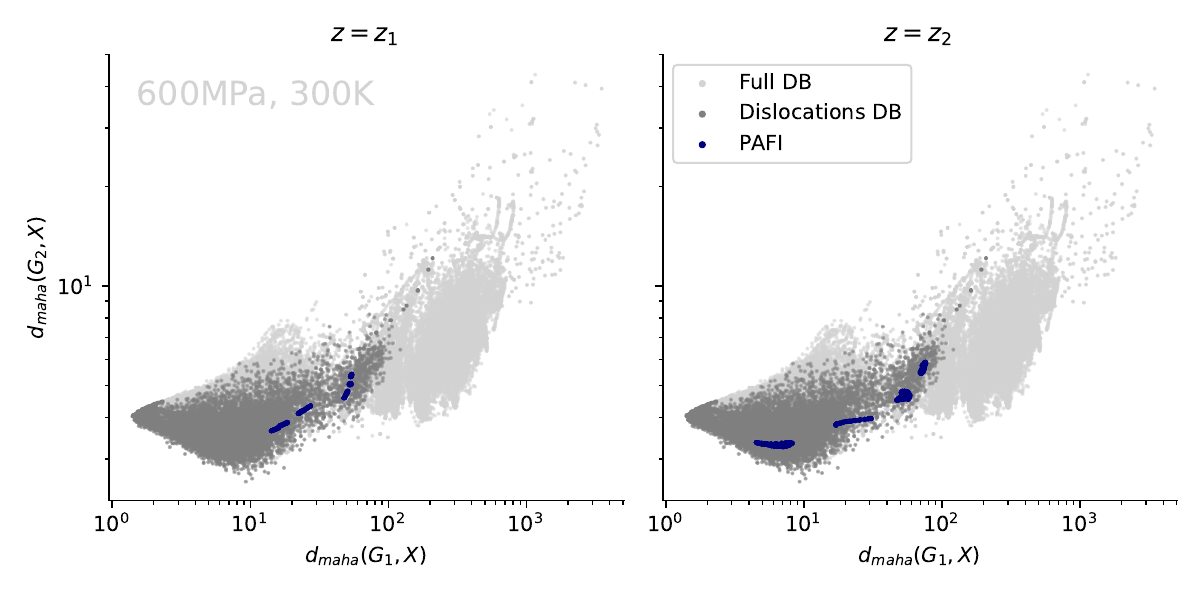}
    \caption{\textbf{Statistical distances of local atomic environments located near dislocation cores discovered by PAFI, compared to the MLIP training database (full database or dislocations configurations only).}
    PAFI configurations are ensemble-averaged atomic positions, extracted from a \SI{600}{\mega\pascal}, \SI{300}{\kelvin} simulation presented in Fig.~2~(e) at two reaction coordinates of interest $z_1$ and $z_2$, which correspond to the minimum and maximum point of the Gibbs energy barrier in these conditions.}
    \label{fig:pafi_maha}
\end{figure}

We also investigate in detail the particularly challenging stress and temperature conditions illustrated in Fig.~2~(e) for the main text, where the coordinate of the minimum Gibbs energy changes towards $z_1$, while the maximum of the profile remains close to $z_2$.
As shown in \ref{fig:pafi_maha}, the sampled domains remain contained in the training database, without out-of-distribution samples.

\FloatBarrier
\section{Additional dislocation properties of the MLIPs}
\label{supp:potentials-properties}

\subsection{Peierls barrier in W}

To validate the new W MLIP, we computed the Peierls barrier associated with the migration of the \textonehalf$[111]$ screw dislocation from one Peierls valley to the next using the NEB method. Similar to Fig.~2(a–b) of the main text, we adopted the simulation setup and elastic correction from Ref.~\cite{bienvenu2022ab} to ensure a direct comparison with DFT.

\begin{figure}[h!tb]
    \centering
    \includegraphics[width=0.5\linewidth]{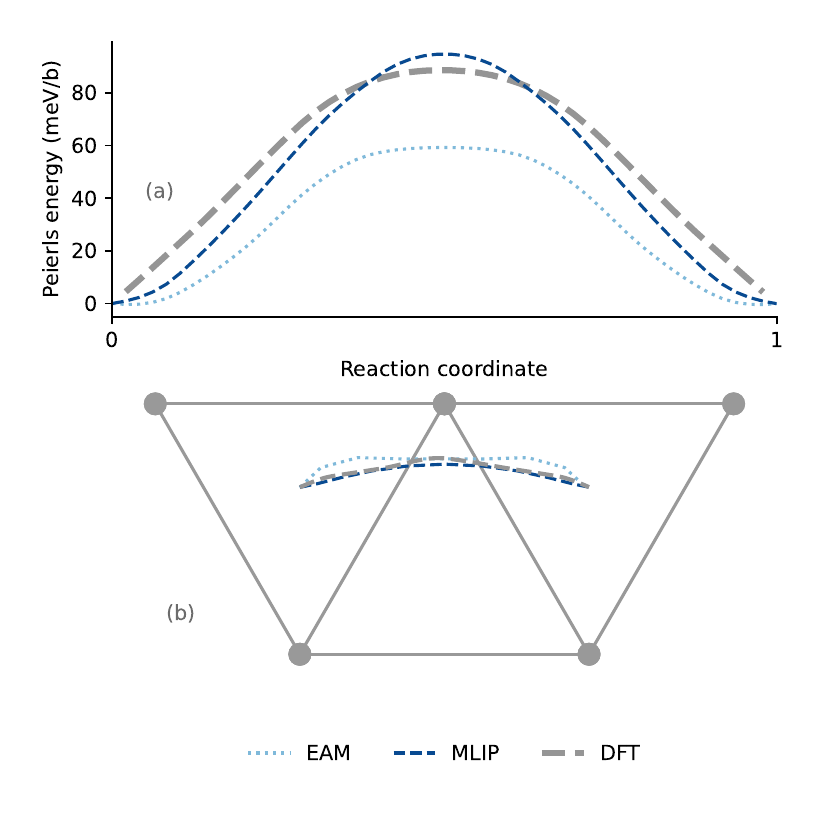}
    \caption{\textbf{Peierls barrier and dislocation core migration trajectory, in W using the EAM and MLIP.} Results are compared to reference DFT calculations from \cite{bienvenu2022ab}.}
    \label{fig:PB-MLIP-W}
\end{figure}

The results are presented in \ref{fig:PB-MLIP-W} and compared to the predictions of the classical EAM potential for W \cite{marinica2013interatomic}, as well as reference DFT calculations from \cite{bienvenu2022ab}.
The new MLIP successfully reproduces the Peierls barrier (\ref{fig:PB-MLIP-W}(a)) with reasonable accuracy relative to DFT while also predicting a core migration trajectory that closely aligns with DFT (\ref{fig:PB-MLIP-W}(b)).

\subsection{Core eigenstrains in Fe and W}

In BCC metals, the yield stress is known to be influenced not only by the resolved shear stress, but also by stress tensor components that do not drive plastic deformation, denoted as non-glide stresses. 
This effect was linked to deformations of the dislocation core along the trajectory between Peierls valleys. These deformations were modeled in  Ref. ~\onlinecite{kraych2019non} using Eshelby inclusions characterized by a core eigenstrain tensor or equivalently a relaxation volume tensor. This property has so far been overlooked when assessing the accuracy of interatomic potentials, despite its major importance on screw dislocation mobility since it directly controls non-glide effects on the yield stress~\cite{kraych2019non,clouet2021screw}. We thus ensured that the improved MLIP predicted accurately the relaxation volume tensor. 

The relaxation volume tensor $\bar{\bar \Omega}$ defined per unit length of dislocation, is expressed along the migration path of the dislocation as:
\begin{equation*}
    \bar{\bar \Omega } = \begin{pmatrix}
                            \Omega_{11} & \Omega_{12} & 0\\
                            \Omega_{12} & \Omega_{22} & 0\\
                            0           &           0 & \Omega_{33}
                          \end{pmatrix}.
\end{equation*}

\begin{figure}[h!tp]
    \centering
    \includegraphics[width=0.5\textwidth]{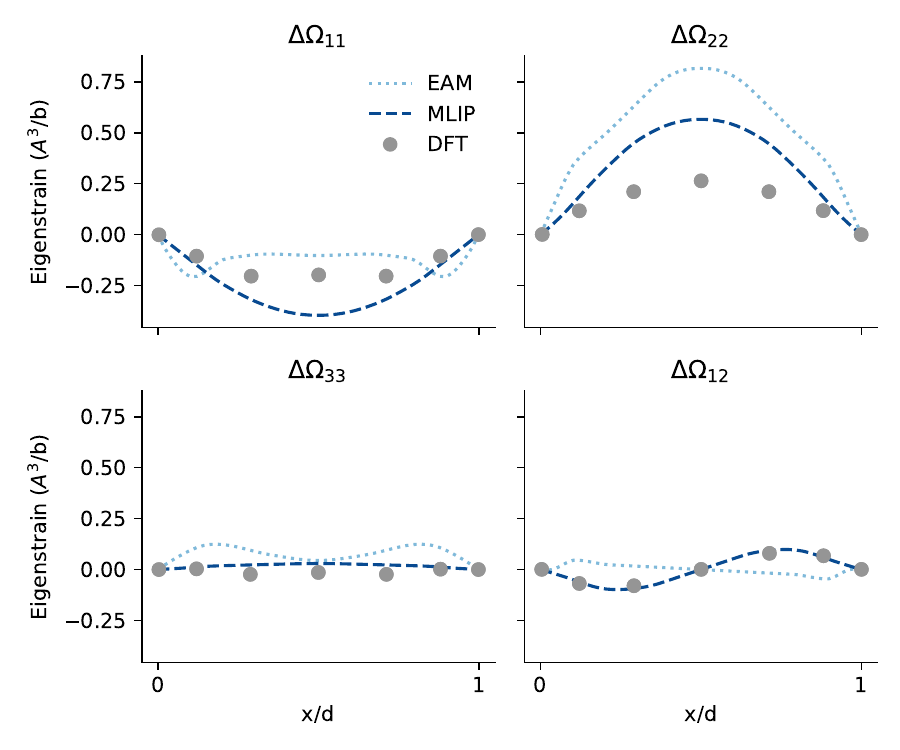}
    \caption{\textbf{Relaxation volumes in Fe}, obtained with the EAM potential and the MLIP for Fe, compared to DFT data from Ref.~\onlinecite{bienvenu2022ab}, using the same methodology. 
    The horizontal axis is the position of the dislocation $X$ divided by the distance between Peierls valleys $d$. 
    }
    \label{fig:eigenstrain}
\end{figure}

The different terms of the tensor can be computed from the variation of the stress tensor along the NEB path of the dislocation between Peierls valleys, as explained in Ref. ~\onlinecite{kraych2019non}. 
The relaxation volumes obtained with the present EAM potential and MLIP for Fe are compared in \ref{fig:eigenstrain} with reference DFT data~\cite{bienvenu2022ab}. With the EAM potential, several relaxation volumes are found in qualitative disagreement with DFT, especially $\Omega_{11}$. By way of contrast, the MLIP predicts smooth evolutions in better agreement with the DFT data.
Combined with the accurate prediction of the core trajectory (see Fig.~2~(b)), we conclude that the present MLIP for Fe accurately accounts for the effect of the full stress tensor components on the glide of the dislocation.

\begin{figure}
    \centering
    \includegraphics[width=0.5\linewidth]{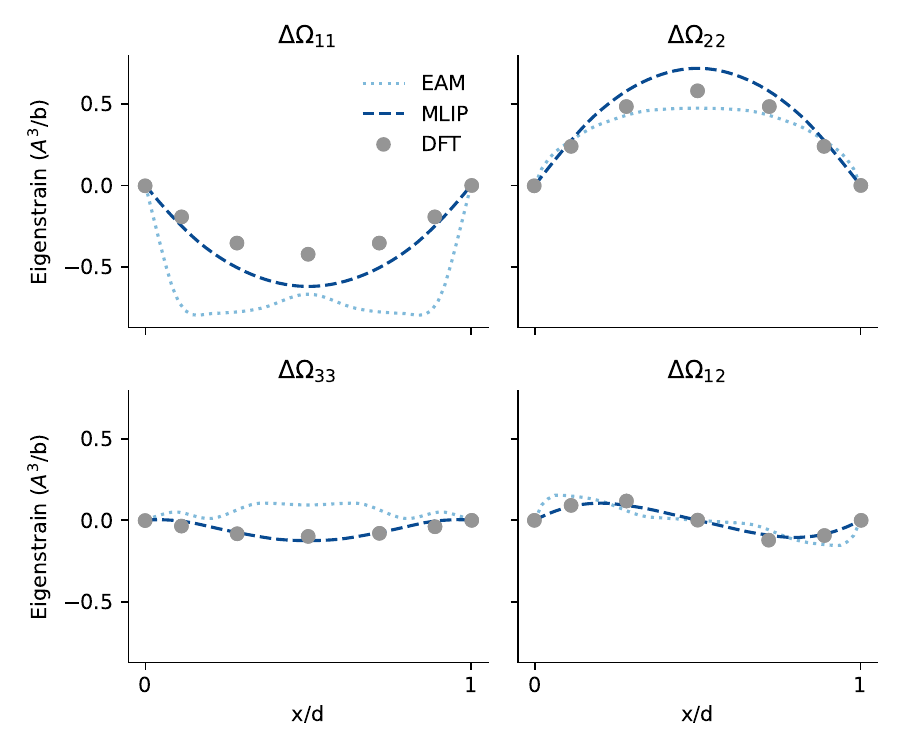}
    \caption{\textbf{Relaxation volumes in W}, obtained with the EAM potential and the MLIP for W, compared to DFT data from Ref.~\onlinecite{bienvenu2022ab}, using the same methodology. 
    The horizontal axis is the position of the dislocation $X$ divided by the distance between Peierls valleys $d$.}
    \label{fig:eigenstrain-W}
\end{figure}

\FloatBarrier
\section*{Supplementary References}
\bibliography{bibliography}%